\documentclass[journal]{IEEEtran}
\usepackage{amsmath,amsthm,amssymb}
\usepackage{mathrsfs,mathtools,mathabx}
\usepackage{bm}
\usepackage{graphicx}
\usepackage{textcomp}
\usepackage{xcolor}
\usepackage{dsfont}
\usepackage{pgfplots}
\pgfplotsset{compat=newest}
\usepackage{subfigure}
\usepackage{cite}
\usepackage{float}
\usepackage{xparse}
\usepackage{hyperref}
\usepackage{tikz}
\usetikzlibrary{calc}
\usetikzlibrary{spy}
\usetikzlibrary{arrows,decorations.pathreplacing,patterns}

\usepackage{etoolbox}

\graphicspath{{./Figures/}}

\newtheorem{prop}{Proposition}
	
\newtheorem{rem}{Remark}

\theoremstyle{definition}

\allowdisplaybreaks 
\usepackage{balance} 
\definecolor{aliceblue}{rgb}{0.74, 0.77, 8.0}
\definecolor{eggplant}{rgb}{0.38, 0.25, 0.32}
\definecolor{pearl}{rgb}{0.94, 0.92, 0.84}
\definecolor{chestnut}{rgb}{0.8, 0.36, 0.36}
\definecolor{airforceblue}{rgb}{0.36, 0.54, 0.66}
\definecolor{cadmiumorange}{rgb}{0.93, 0.53, 0.18}
\definecolor{bleudefrance}{rgb}{0.19, 0.55, 0.91}
\definecolor{carolinablue}{rgb}{0.6, 0.73, 0.89}
\definecolor{blue(ncs)}{rgb}{0.0, 0.53, 0.74}
\definecolor{dodgerblue}{rgb}{0.12, 0.56, 1.0}
\definecolor{cssgreen}{rgb}{0.0, 0.5, 0.0}
\definecolor{cadmiumgreen}{rgb}{0.0, 0.42, 0.24}
\definecolor{cadmiumorange}{rgb}{0.93, 0.53, 0.18}
\definecolor{amaranth}{rgb}{0.9, 0.17, 0.31}
\definecolor{bluegray}{rgb}{0.4, 0.6, 0.8}
\definecolor{cadmiumgreen}{rgb}{0.0, 0.42, 0.24}
\definecolor{amaranth}{rgb}{0.9, 0.17, 0.31}
\definecolor{amethyst}{rgb}{0.6, 0.4, 0.8}
\definecolor{amber}{rgb}{1.0, 0.75, 0.0}
\definecolor{azure}{rgb}{0.0, 0.5, 1.0}
\definecolor{babyblue}{rgb}{0.54, 0.81, 0.94}
\definecolor{bazaar}{rgb}{0.6, 0.47, 0.48}
\definecolor{celestialblue}{rgb}{0.29, 0.59, 0.82}
\definecolor{darklavender}{rgb}{0.45, 0.31, 0.59}
\definecolor{bluebell}{rgb}{0.64, 0.64, 0.82}
\definecolor{chamoisee}{rgb}{0.63, 0.47, 0.35}
\definecolor{darkcerulean}{rgb}{0.03, 0.27, 0.49}
\definecolor{iris}{rgb}{0.35, 0.31, 0.81}
\definecolor{jazzberryjam}{rgb}{0.65, 0.04, 0.37}

\definecolor{egyptianblue}{rgb}{0.06, 0.2, 0.65}
\definecolor{brown(web)}{rgb}{0.65, 0.16, 0.16}
\definecolor{burntumber}{rgb}{0.54, 0.2, 0.14}
	\definecolor{charcoal}{rgb}{0.21, 0.27, 0.31}
\definecolor{camel}{rgb}{0.76, 0.6, 0.42}
\definecolor{chamoisee}{rgb}{0.63, 0.47, 0.35}
	\definecolor{darkcerulean}{rgb}{0.03, 0.27, 0.49}

\begin{document}

\title{On Designing Modulation for Over-the-Air Computation — Part I: Noise-Aware Design}

\author{ Saeed Razavikia,~\IEEEmembership{Member,~IEEE}, Carlo Fischione,~\IEEEmembership{Fellow,~IEEE}
\thanks{S. Razavikia and C. Fischione are with the School of Electrical Engineering and Computer Science, KTH Royal Institute of Technology, Stockholm, Sweden (e-mail: \{sraz, carlofi\}@kth.se). C. Fischione is also with Digital Futures of KTH. 

}
\thanks{S. Razavikia was jointly supported by the Wallenberg AI, Autonomous Systems and Software Program (WASP) and the Ericsson Research Foundation. The  SSF SAICOM project, the Digital Futures project DEMOCRITUS, and the Swedish Research Council Project MALEN partially supported this work.}
}

\maketitle

\begin{abstract}
Over-the-air computation (OAC) leverages the physical superposition property of wireless multiple access channels (MACs) to compute functions while communication occurs, enabling scalable and low-latency processing in distributed networks. While analog OAC methods suffer from noise sensitivity and hardware constraints, existing digital approaches are often limited in design complexity, which may hinder scalability and fail to exploit spectral efficiency fully. This two-part paper revisits and extends the ChannelComp framework, a general methodology for computing arbitrary finite-valued functions using digital modulation. In Part I,  we develop a novel constellation design approach that is aware of the noise distribution and formulates the encoder design as a max-min optimization problem using noise-tailored distance metrics. Our design supports noise models, including Gaussian, Laplace, and heavy-tailed distributions. We further demonstrate that, for heavy-tailed noise, the optimal ChannelComp setup coincides with the solution to the corresponding max–min criterion for the channel noise with heavy-tailed distributions. Numerical experiments confirm that our noise-aware design achieves a substantially lower mean-square error than leading digital OAC methods over noisy MACs. In Part II, we consider a constellation design with a quantization-based sampling scheme to enhance modulation scalability and computational accuracy for large-scale digital OAC.
\end{abstract}


\section{Introduction}
The main idea of over-the-air computation (OAC) is to leverage the superposition property of wireless multiple access channels to compute mathematical functions~\cite{nazer2007Computation}. In a communication network, standard communication protocols require the nodes to transmit data over an orthogonal channel to avoid interference. Instead,  OAC offers the computation by harnessing the interference via simultaneous transmissions~\cite{csahin2023survey}. To be concrete, suppose that the goal is to evaluate a function $f(s_1,\ldots,s_K)$ at the server or computation point, connected to $K$ nodes owning the symbol $s_k$s values.  In standard communication protocols, each node needs to send the real value $s_k$ over individual communication resources such as time slots and frequency. However, OAC 
aims to compute the function via signal superposition in the channel. In this regard, the network can significantly reduce communication resources as the number of nodes in the network $K$ grows.   

Such a disruptive concept can potentially reform the traditional way of independently handling computation and communication tasks. Indeed, considering the rapid growth rate of large language models (LLMs)\cite{xue2023repeat} and scaling up transformer-based language models~\cite{kaplan2020scaling}, the demand for more power-efficient protocols for both communication and computation becomes more apparent.  For instance, for training a reasonably small LLM  BLOOM~\cite{ai2022bigscience}, emitted approximately $50.5$ tonnes of $\text{CO}_2$~\cite{luccioni2023estimating}. On the other hand, due to the massive size of such models with trillions of parameters, utilizing a distributed computing platform for the training, fine-tuning, and inference is unavoidable \cite{zeng2023distributed}. Therefore, OAC could potentially improve energy 
efficiency and accelerate the computation for such distributed computing systems ~\cite{chen2023over,mitsiou2023accelerating}. Moreover,  this technique can be used to scale up radio and computation resources for compute-intensive applications over distributed networks, such as distributed learning~\cite{hellstrom2022wireless,amiri2020federated}, wireless control system~\cite{cai2017modulation}, and wireless intra-chip computation~\cite{guirado2023whype}.

In this two-part paper, we consider designing modulation for the OAC problem to improve the scheme's computation capability and communication reliability. Specifically, in Part I, we propose novel criteria for designing digital modulation for OAC over a multiple access channel (MAC).  We develop our scheme for different channel noise distributions such as white Gaussian, Laplace, and heavy tail. Furthermore, we extend the scheme for stochastic fading channels.  In Part II, we study the sampling and quantization problem for digital OAC and show how suitable sampling can considerably reduce complexity while improving computation reliability.

\subsection{Literature Review}

Early analyses on computing linear functions over multiple access channels were studied in \cite{nazer2007Computation,nazer2011compute} and later extended to compute nomographic functions in wireless sensor networks as shown in~\cite{Golden2013Harnessing,goldenbaum2014nomographic}. However, since these OAC techniques primarily relied on analog amplitude modulation, they did not offer a viable solution for integration with existing wireless hardware \cite{chen2018over,ang2019robust,goldenbaum2013robust}.  Additionally, the inherent vulnerability of analog modulation to channel impairments—such as noise and fading—detracts from its communication reliability \cite{csahin2022over}. As a result, digital modulation has garnered preference, attributed to its superior channel correction performance and widespread integration~\cite{razavikia2023computing}. In the pursuit of incorporating digital modulation into OAC systems, research has explored the application of simple digital schemes, such as BPSK and FSK, tailored for specific machine learning training operations over wireless networks~\cite{bernstein2018signsgd,csahin2022over,csahin2023distributed,csahin2024over}. Despite these advancements, the resultant methods often exhibit inefficient utilization of resources~\cite{per2024waveforms}.

To overcome these limitations, our prior works~\cite{razavikia2023computing, saeed2023ChannelComp, razavikia2023SumCode, razavikia2024FedComp} introduced \textit{ChannelComp}, a general digital framework for computing any finite-valued function over the MAC. The core idea is to design modulations that avoid destructive \textit{constellation points overlaps}, enabling reliable and efficient computation over noisy wireless channels.  In \cite{Yan2024Novel,yan2025remac}, the authors extend the idea over multiple time slots to improve the communication reliability scheme.  However, ChannelComp’s framework requires high computational complexity to design modulation whenever the modulation order or the number of nodes is exceedingly high. To overcome such complexity, in \cite{li2025channel}, a channel-aware constellation design scheme or demodulation mapper is applied to cellular and cell-free massive MIMO systems, where they dynamically adapt the constellation based on channel conditions to reduce computational complexity. Although such constellation design schemes avoid solving an optimization, they result in performance reduction and inefficient utilization of resources.

In classical digital communication, constellation design has long been a fundamental problem~\cite{forney1998modulation,makowski2006optimality,zhang2019constellation}, with cubic, circular, and rectangular geometries widely adopted in standards~\cite{cioffi2014signal}. These designs are typically shaped to optimize error probability, spectral efficiency, or power constraints~\cite{sung2009performance}. Depending on the noise characteristics, the design may favor dense packing for capacity (e.g., equiprobable signaling under Gaussian noise~\cite{sun1993approaching}) or adapt to distortion effects, such as phase noise or non-Gaussian interference~\cite{kayhan2014constellation}.

Design criteria generally aim to minimize symbol error rates~\cite{foschini2003optimization} or maximize mutual information. In particular, warping techniques have been shown to be effective at high variance of the noise to approach capacity, while uniformly spaced constellations are more suitable in low variance of the channel noise (high signal-to-noise ratio) regimes~\cite{betts1994performance}. Additionally, channel-aware designs incorporating transmitter-side channel state information have emerged in spatial modulation contexts~\cite{maleki2012space}.

In our context of digital OAC, these classical geometrical insights motivate designing modulation schemes that account for both the desired function's structure and the channel's statistical nature. Unlike classical systems where minimizing bit errors is sufficient, in OAC, the constellation must also avoid overlaps in the superimposed signal domain, which directly translates to computation ambiguity~\cite{saeed2023ChannelComp}. Thus, we reinterpret constellation design as a noise-aware optimization problem over the complex field, guided by function computation and the underlying channel distribution.

\begin{figure*}
    \centering
\definecolor{palecgray}{rgb}{0.93, 0.94, 0.94}
\tikzset{every picture/.style={line width=0.75pt}} 

\begin{tikzpicture}[x=0.75pt,y=0.75pt,yscale=-1,xscale=1]


\begin{scope}[shift={(-1.2cm,0)}]

\draw [dashed, color=bazaar]   (383pt,70pt) -- (383pt,165pt) ;
\draw [dashed, color=bazaar]   (433pt,70pt) -- (433pt,165pt) ;

\draw [dashed, color=bazaar]   (485pt,70pt) -- (485pt,165pt) ;

\draw[fill=palecgray , rounded corners=5pt] (470pt, 130pt) rectangle (440pt, 100pt) {};
\draw (455pt,115pt) node   {$\mathcal{T}(\cdot)$};

\draw [-latex]    (420pt,115pt) -- (440pt,115pt) ;

\draw[fill=palecgray , rounded corners=5pt] (520pt, 130pt) rectangle (490pt, 100pt) {};
\draw (505pt,115pt) node   {$\mathcal{Q}_p(\cdot)$};

\draw [-latex]    (520pt,115pt) -- (540pt,115pt) ;

\draw[fill=palecgray , rounded corners=5pt] (420pt, 130pt) rectangle (390pt, 100pt) {};
\draw (405pt,115pt) node   {$\mathcal{D}(\cdot)$};
\draw [-latex]    (370pt,115pt) -- (390pt,115pt) ;

\draw (540,80) node {\scriptsize Modulation  Decoding};
\draw (620,80) node {\scriptsize Tabular Mapper };
\draw (680,80) node {\scriptsize Sampling};
\draw [-latex]  (310pt,80pt) -- (354pt,110pt) ;
\draw [-latex]  (310pt,150pt) -- (354pt,120pt) ;

\draw (360pt, 115pt) node {\LARGE $\bigoplus$};
\draw (480,105) node {\color{brown(web)}$z$};

\draw [-latex]  (480,85pt) -- (480,105pt) ;

\draw (500,145) node {$r$};
\draw (570,145) node {$\tilde{x}$};

\draw [-latex]  (470pt,115pt) -- (490pt,115pt) ;
\draw (635,143) node {$\hat{f}$};

\draw (710,143) node {$\tilde{f}$};

\draw  [color={rgb, 255:red, 74; green, 144; blue, 226 }][dash pattern={on 0.84pt off 2.51pt}] (505,125) -- (700,125) -- (700,180) -- (505,180) -- cycle ;

\draw (600,115) node {\color{rgb, 255:red, 74; green, 144; blue, 226 } $\mathscr{D}$};

\end{scope}

\draw[-latex]   (200pt,80pt) -- (220pt,80pt) ;

\draw (227pt, 50pt) node { $p_1$};
\draw [-latex]  (227pt,55pt) -- (227pt,75pt) ;
\draw (227pt, 80pt) node { $\bigotimes$};

\draw  [-latex]  (235pt,80pt) -- (255pt,80pt) ;

\draw [-latex]  (227pt,125pt) -- (227pt,145pt) ;
\draw (227pt, 150pt) node { $\bigotimes$};
\draw (227pt, 120pt) node { $p_K$};
\draw [-latex]   (235pt,150pt) -- (255pt,150pt) ;

\draw  [color={rgb, 255:red, 74; green, 144; blue, 226 }][dash pattern={on 0.84pt off 2.51pt}] (85,80) -- (275,80) -- (275,130) -- (85,130) -- cycle ;

\draw[fill=palecgray , rounded corners=5pt] (200pt, 95pt) rectangle (170pt, 65pt) {};
\draw (185pt,80pt) node   {\color{brown(web)}$\mathcal{E}(\cdot)$};

\draw [-latex]    (150pt,80pt) -- (170pt,80pt) ;

\draw[fill=palecgray , rounded corners=5pt] (150pt, 95pt) rectangle (120pt, 65pt) {};
\draw (135pt,80pt) node   {$\mathcal{Q}(\cdot)$};
\draw [-latex]    (100pt,80pt) -- (120pt,80pt) ;

\draw[fill=palecgray , rounded corners=5pt] (100pt, 95pt) rectangle (70pt, 65pt) {};
\draw (85pt,80pt) node   {$\varphi_1(\cdot)$};
\draw [-latex]    (50pt,80pt) -- (70pt,80pt) ;

\draw[-latex]    (200pt,150pt) -- (220pt,150pt) ;

\draw[fill=palecgray , rounded corners=5pt] (200pt, 165pt) rectangle (170pt, 135pt) {};
\draw (185pt,150pt) node   {\color{brown(web)}$\mathcal{E}(\cdot)$};

\draw [-latex]    (150pt,150pt) -- (170pt,150pt) ;

\draw[fill=palecgray , rounded corners=5pt] (150pt, 165pt) rectangle (120pt, 135pt) {};
\draw (135pt,150pt) node   {$\mathcal{Q}(\cdot)$};
\draw [-latex]    (100pt,150pt) -- (120pt,150pt) ;

\draw[fill=palecgray , rounded corners=5pt] (100pt, 165pt) rectangle (70pt, 135pt) {};
\draw (85pt,150pt) node   {$\varphi_K(\cdot)$};
\draw [-latex]    (50pt,150pt) -- (70pt,150pt) ;


\draw  [color={rgb, 255:red, 74; green, 144; blue, 226 }][dash pattern={on 0.84pt off 2.51pt}] (85,175) -- (275,175) -- (275,225) -- (85,225) -- cycle ;

\draw [dashed, color=bazaar]   (114pt,40pt) -- (114pt,175pt) ;
\draw [dashed, color=bazaar]   (164pt,40pt) -- (164pt,175pt) ;

\draw (180,70) node {\color{rgb, 255:red, 74; green, 144; blue, 226 } $\mathscr{E}_1$};

\draw (105,40) node {\scriptsize Source Encoding};
\draw (185,40) node {\scriptsize Quantization};
\draw (270,40) node {\scriptsize Modulation Encoding};

\draw (75,100) node {$s_1$};
\draw (145,100) node {$c_1$};
\draw (210,100) node {$\tilde{c}_1$};
\draw (289,100) node {$x_1$};
\draw (350,100) node {$x_1p_1$};
\draw (400,110) node {$h_1$};

\draw (75,145) node {\Large $\vdots$};
\draw (145,145) node {\Large $\vdots$};
\draw (210,145) node {\Large $\vdots$};
\draw (280,145) node {\Large $\vdots$};
\draw (380,145) node {\Large $\vdots$};

\draw (75,190) node {$s_K$};
\draw (145,190) node {$c_K$};
\draw (210,190) node {$\tilde{c}_K$};
\draw (289,190) node {$x_K$};
\draw (350,190) node {$x_Kp_K$};
\draw (400,200) node {$h_K$};

\draw (180,160) node {\color{rgb, 255:red, 74; green, 144; blue, 226 } $\mathscr{E}_K$};

\end{tikzpicture}

\caption{Block diagram of the communication model. Each node $k$ encodes its value $s_k$ via source encoding $\varphi_k(\cdot)$, quantization $\mathcal{Q}(\cdot)$, and modulation $\mathcal{E}(\cdot)$ to generate $x_k$. The signals are weighted by $p_k$ and transmitted over the MAC with channel coefficients $h_k$. At the CP, the received signal $r$ is processed by a decoder $\mathcal{D}(\cdot)$ and a tabular mapper $\mathcal{T}(\cdot)$ to estimate the function output $\hat{f}$. Finally, the estimated output $\hat{f}$ is quantized (sampled) by $\mathcal{Q}_p(\cdot)$ to yield $\tilde{f}\in \mathcal{Y}_f^{p}$, where $p$ denotes order of quantization. The encoding operators $\mathscr{E}_k$ and the decoding operator $\mathscr{D}$ enable efficient computation via communication. In Part I, we focus on designing the encoder operator {\color{brown(web)} $\mathcal{E}(\cdot)$ }while accounting for the distribution of the channel noise {\color{brown(web)} $z$} where are depicted by color code {\color{brown(web)} brown}. }

    \label{fig:Systemmodel}
\end{figure*}
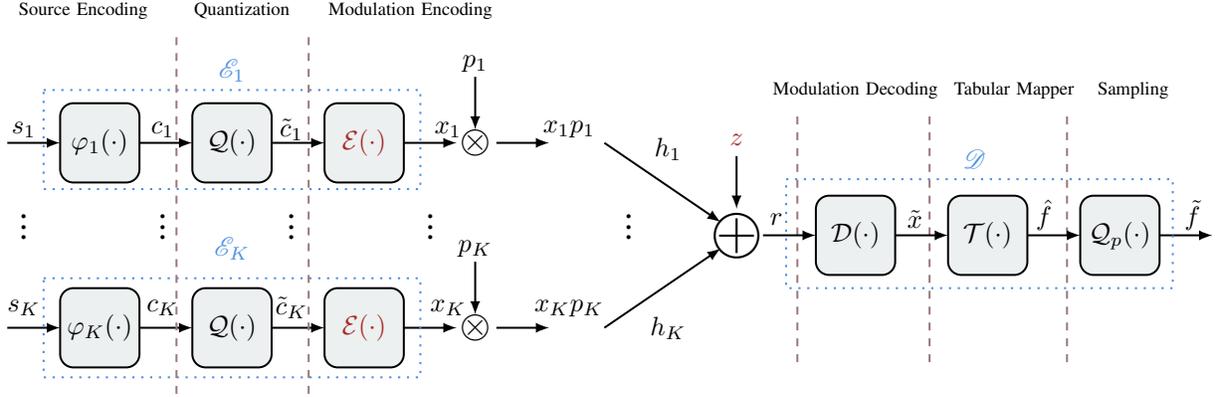

\subsection{Key Contributions of Part I}

In this work, we go beyond traditional constellation design by introducing a computation-oriented scheme for designing modulations. To this end, we introduce a novel distance-based optimization framework that aligns constellation spacing with the tail behavior of the channel noise distribution, which directly governs the likelihood of computational errors. This enables our modulation design to dynamically adapt to channel noise distributions, such as Gaussian, Laplace, and heavy-tailed (e.g., Cauchy), by selecting appropriate surrogate distance metrics, including exponential, polynomial, or power-law forms.

Moreover, we show that minimizing the maximum computation error over the MAC is approximately equivalent to reducing the mean square computation error (MSE) under sub-Gaussian noise in the low-variance regime. This insight justifies using a max-min optimization approach, which proves effective across various statistical noise models. Notably, our perspective reframes constellation design by ChannelComp~\cite{saeed2023ChannelComp} as a distribution-aware function encoding problem, where the modulation is jointly optimized to the channel statistics functional structure. This extends the ChannelComp framework and enables scalable, robust digital OAC schemes suitable for modern distributed systems.

In short, our contributions can be listed below. 

\begin{itemize} 
\item \textbf{Noise-aware modulation design:} We formulate the modulation design as a max-min optimization problem that minimizes the maximum computation error over a noisy MAC. This optimization naturally leads to constellation structures that respect the geometry of the function and the channel noise distributions.

\item \textbf{Design under various noise distributions:} We extend the framework to support a broad family of noise models, including Gaussian, Laplace (generalized normal), and heavy-tailed (e.g., Cauchy) distributions. Each model leads to a distinct distance metric, shaping the constellation accordingly. Notably, we show that the standard ChannelComp distance choice in \cite{saeed2023ChannelComp} is a special case for the channel noise with a heavy-tailed distribution.

\item \textbf{MSE and minimizing the maximum computation error:} We show that for the channel noise with sub-Gaussian distributions, when the variance is small, minimizing mean square computation error is approximately equivalent to reducing the maximum computation error, thus justifying the use of a max-min framework for the low variance regimes.

\item \textbf{Numerical Results:} Through simulations on sum and product functions, we show that our noise-aware modulation scheme significantly outperforms baseline methods, across various channel conditions and distributions. 
\end{itemize}

In both parts of this work, we revisit the ChannelComp system model and adopt a slightly modified communication architecture to better suit practical implementation by considering sampling and quantization. These aspects are crucial for scaling the framework to real-world applications and are addressed in detail in Part~II~\cite{razavi2025revisitII}. 

\subsection{Organization of the Paper}

 Section~\ref{sec:system_model} presents the system model and defines our target class of functions for performing the computation. Section~\ref{sec:channelComp} provides the preliminaries on the ChannelComp encoder-decoder principle and discusses constellation overlap avoidance. Section~\ref{sec:DesingCriteria} provides the noise-aware modulation design criterion for different noise distributions. Then, Section~\ref{sec:Evaluation} presents the simulation results and evaluates the performance of the proposed design criterion. Finally, Section~\ref{sec:Conclusion} concludes the paper.

\subsection{Notation}

We denote by $\mathbb{Z}$, $\mathbb{R}$, and $\mathbb{C}$ the sets of integers, real numbers, and complex numbers, respectively. Scalars are represented by lowercase letters such as $x$, while operators are denoted using calligraphic letters like $\mathcal{A}$. For a complex number $x \in \mathbb{C}$, its real and imaginary parts are denoted by $\mathfrak{Re}(x)$ and $\mathfrak{Im}(x)$, respectively. The Euclidean norm of a vector $\bm{x} \in \mathbb{C}^n$ is denoted by $\|\bm{x}\|_2$, and its squared form is $\|\bm{x}\|_2^2 = \sum_i |x_i|^2$. For a set $\Omega$, we use $|\Omega|$ to present the cardinality of the set. For integers $n,k\in\mathbb{Z}^{+}$ with $0\le k\le n$, the binomial coefficient is denoted by $\binom{n}{k}=n!/(n-k)!k!$.  For an integer $N$, $[N]$ corresponds to the set  $\{1,2,\dots, N\}$.

The distribution $\mathcal{CN}(0,\sigma^2)$ denotes a circularly symmetric complex Gaussian distribution with independent real and imaginary components, each following a normal distribution $\mathcal{N}(0,\sigma^2/2)$.

\section{System Model}\label{sec:system_model}

In this section, we introduce the distributed network system and the communication model of the network, and define the class of functions that our scheme can efficiently compute over the network. Specifically, we consider a network where multiple nodes transmit modulated signals over a shared medium, leveraging the principles of the OAC.

\subsection{Which Class of Functions?}

We note that while the proposed scheme in this paper can be applied to any many-to-one function in the form of $f(s_1,\ldots,s_K)$, for the sake of computational complexity that is studied in Part~II~\cite{razavi2025revisitII},  we narrow down our target functions to a set of functions called \textit{aggregation functions}. Let $s_k\in \mathbb{R}$ for $K\geq 2$,   and $g:\mathbb{R}^{K}\mapsto \mathbb{R}$ be a symmetric  functions, i.e., 
\begin{align}
  \label{eq:sym}
  g(s_1,\ldots,s_K) = g(\upsilon(s_1),\ldots,\upsilon(s_K)),  
\end{align}
for all possible permutations by $\upsilon: \{1,\ldots,K\}\mapsto \{1,\ldots,K\}$. Then, we consider the set of aggregation functions $f: \mathbb{R}^{K}\mapsto \mathbb{R}$, which has the following form. 
\begin{align}
     \label{eq:aggregations}
    f =  g(\phi_1(s_1),\ldots,\phi_K(s_K)),
\end{align}
 where $\phi_k(s_k): \mathbb{R} \mapsto \mathbb{R}$. We note that the presented form in~\eqref{eq:aggregations} is more general than the class of functions that AirComp can handle (nomographic functions). Some examples of these aggregation functions are listed below:
 \begin{itemize}
     \item For $g(y_1,\ldots,y_K) := \psi (\sum_{k=1}^Ky_k)$, the aggregation function becomes nomographic functions. 
     \item  For $g(y_1,\ldots,y_K):= \psi (\prod_{k=1}^Ky_k)$, the aggregation function becomes a multiplicative nomographic function, where an essential feature of the symmetric function is to present the identity privacy of the nodes. 
     \item If $\phi_k$ becomes identity map, i.e., $\phi_k(s)=s$ $\forall~k$, then $f$ become general symmetry functions. 
     \item For  $g(y_1,\ldots,y_K) = \psi\Big(\max_{k\in[K]} \{y_k\}\Big)$, the aggregation function represents a max-aggregation function, which is particularly useful in applications such as anomaly detection and distributed decision making.
 \end{itemize}

These functions are particularly relevant in applications such as sensor networks~\cite{goldenbaum2014nomographic} and distributed learning~\cite{amiri2020federated}, where aggregating distributed data efficiently is essential.

\subsection{Communication Model}

\input{Figs/Fig_BPSK}

Here,  we consider a typical communication model for the OAC problem,  where a network of $K$ nodes and a computational point (CP) is connected through a common communication channel.  To be able to compute the function in the form of \eqref{eq:aggregations}, node $k$ first encodes its value $s_k\in \mathbb{R}$ to $c_k\in \mathbb{R}$ using the source encoder $\varphi_k$, i.e., $c_k:=\varphi_k(s_k)$. Then, $c_k$ is quantized with $q$ levels by a quantizer operator $\mathcal{Q}$, i.e., $\tilde{c}_k =\mathcal{Q} (c_k) \in \mathcal{X}_k$ with $|\mathcal{X}_k|\leq q$. Then, the quantized value $\tilde{c}_k$ is modulated by using encoder $\mathcal{E}(\cdot):\mathcal{X}_k\mapsto \mathbb{C}$ to create a digitally modulated signal ${x}_k$\footnote{Since the function $g$ is a symmetry function, it is enough to use the same modulation encoder $\mathcal{E}$ for all the nodes~\cite{razavikia2023computing}. }. The modulated signal  ${x}_k \in \mathbb{C}$ is a complex-valued signal whose real and imaginary elements correspond to in-phase and quadrature components. Without loss of generality, we assume the modulated signals are normalized such that $|x_k|^2\leq 1$, for $k\in [K]$.  Then, each node $k$ transmits ${x}_k$ with transmission power $p_k$ simultaneously\footnote{ Imperfect synchronization among the nodes can be handled by using the existing techniques,  e.g., \cite{saeed2022BlindFed,hellstrom2023optimal,daei2025timely}.} transmit over the MAC, where the CP server obtains the sum of all $x_k$'s, i.e., 
\begin{align}
  \label{eq:channel}
            r = \sum_{k=1}^{K}p_kh_kx_k + {z}, 
\end{align}
where $r$ is the received signal through the superposition of electromagnetic waves, $h_k$ indicates the channel coefficient between node $k$ and the CP. Also,  $z$ is Gaussian noise with zero mean and variance $\sigma^2$, i.e.,  $z\sim \mathcal{N}(0,\sigma^2)$. To compensate for the fading effects in \eqref{eq:channel}, we use channel inversion power control~\cite{zhu2020one,zhu2019broadband}, where we adjust the transmit power as the inverse of the channel, i.e., $p_k = h_k^{*}/|h_k|^2$ for $k\in [K]$.  Let $\mathscr{E}_k$ be the encoder operator for the encoding procedure at node $k$; we can rewrite Eq.~\eqref{eq:channel} as
\begin{align}
  \label{eq:channelfree}
            r = \sum_{k=1}^{K}\mathscr{E}_k(s_k) + \tilde{z}, 
\end{align}
where $\tilde{z}$ includes the noise resulting from the imperfect channel compensation and the receiving noise.  Next, the CP uses decoder  $\mathcal{D}$ to estimate the transmitted value $\tilde{x}:= \sum_kx_k$. Afterward, by applying a tabular mapper, the desired function $f$ can be computed, i.e., $\hat{f}:= \mathcal{T}(\tilde{x})$, where $\mathcal{T}: \mathbb{C}\mapsto \mathcal{Y}_f$ maps the estimated constellation diagram to the output of the desired function $f\in \mathcal{Y}_f$, where $\mathcal{Y}_f$ denotes the set of all possible values with quantized input domain. Finally, the estimated value $\hat{f}$ is quantized by the sampler operator $\mathcal{Q}_p$, to obtain the $\tilde{f} = \mathcal{Q}_p(\hat{f})$.  The decoding process $\mathscr{D}$ consists of two main components: 
\begin{itemize}
    \item \textit{Estimator}:  Given proper encoders $\mathcal{E}(\cdot)$, the optimal decoder $\mathcal{D}(\cdot)$ can be obtained by maximum likelihood or maximum posterior criteria \cite{saeed2023ChannelComp,liu2024digital}.  Here, we consider the maximum likelihood decoder as in \cite{saeed2023ChannelComp} for simplicity and to keep the generality without imposing any distribution on the input values. 
    \item \textit{Tabular map}: The tabular mapper $\mathcal{T}(\cdot)$ is dictated  by the function output and the encoder operator $\mathcal{E}(\cdot)$. However, since here $\tilde{f}$ is a quantized value, we will show that $\mathcal{T}(\cdot)$ can play a crucial role in reducing the complexity of designing $\mathcal{E}(\cdot)$ by sampling a subset of the function outputs, thereby enabling a more efficient computation process. In Part~II~\cite{razavi2025revisitII}, we focus on designing such a sampling strategy. 
\end{itemize}

 The overall communication model is depicted in Figure~\ref{fig:Systemmodel}. In this paper, the main goal is to devise proper criteria for designing the encoders $\mathcal{E}(\cdot)$ to efficiently compute the desired function $f$ at the CP by taking into account the distribution of the noise $z$~\eqref{eq:channelfree}. 
 In the next section, we discuss the challenges and criteria in designing the decoders $\mathscr{E}_k(\cdot)$'s and how they can be addressed.  

\begin{figure*}
    \centering

    \subfigure[Superposition of two PAM-$4$]{

\begin{tikzpicture}
    \foreach \x in {0,1,2,3,4,5,6} {
      \node[circle, fill=black, inner sep=1.5pt] at (\x,0)   {};
      \foreach \i in {1,...,30} {
           \pgfmathsetmacro{\Uone}{rnd}
           \pgfmathsetmacro{\Utwo}{rnd}
           \pgfmathsetmacro{\noiseX}{0.2*sqrt(-2*ln(\Uone))*cos(360*\Utwo)}
           \pgfmathsetmacro{\noiseY}{0.2*sqrt(-2*ln(\Uone))*sin(360*\Utwo)}
           \node[scale=0.5] at ({\x+\noiseX}, {0+\noiseY}) {\color{chamoisee}$\circ$};
      }
      \node at (\x,10pt) {\footnotesize$\x$};
    }
   \draw[<->]  (1.15,0) -- (1.85,0);
   \node  at (1.5,10pt) {\footnotesize $d$};
   \node  at (2.5,10pt) {\footnotesize $d$};
   \node  at (2,30pt) {\footnotesize $2d$};
\draw[<->]  (2.15,0) -- (2.85,0);
   \draw[<->]  (1.15,0.8) -- (2.85,0.8);
 \end{tikzpicture}
  }
  \subfigure[Superposition of two skewed QAM-$4$]{
\begin{tikzpicture}

    \foreach \x/\y in {0/0, 1.732/0, 3.464/0, 0.5/1, 2.232/1, 3.964/1, 1/2, 2.732/2, 4.464/2} {
      \node[circle, fill=black, inner sep=1.5pt] at (\x,\y) {};
      \foreach \i in {1,...,30} {
           \pgfmathsetmacro{\Uone}{rnd}
           \pgfmathsetmacro{\Utwo}{rnd}
           \pgfmathsetmacro{\noiseX}{0.2*sqrt(-2*ln(\Uone))*cos(360*\Utwo)}
           \pgfmathsetmacro{\noiseY}{0.2*sqrt(-2*ln(\Uone))*sin(360*\Utwo)}
           \node[scale=0.5] at ({\x+\noiseX}, {\y+\noiseY}) {\color{chamoisee}$\circ$};
      }
    }  
   \node at (0,0.4) {\footnotesize$ 0$};
   \node at (1.732,0.4) {\footnotesize$ 2$};
   \node at (3.464,0.4) {\footnotesize$ 4$};
   
   \node at (0.5,1.4) {\footnotesize$ 1$};
   \node at (2.232,1.4) {\footnotesize$ 3$};
   \node at (3.964,1.4) {\footnotesize$ 5$};
   
   \node at (1,2.4) {\footnotesize$ 2$};
   \node at (2.732,2.4) {\footnotesize$ 4$};
   \node at (4.464,2.4) {\footnotesize$ 6$};

   \draw[<->]  (0.6,1.1) -- (1, 1.9);
   \node at (0.7,1.7) {\footnotesize$d$};
   \draw[<->]  (1.1, 1.9) -- (2.2,1.2);
   \node at (1.8,1.85) {\footnotesize$ \sqrt{\tfrac{5}{2}}d$};

   \draw[<->]  (0.7, 1) -- (2, 1);
   \node at (1.3,0.75) {\footnotesize$\sqrt{3}d$};
\end{tikzpicture}
    }
    \caption{Superimposed constellation diagrams for two modulation formats. Subfigure (a) shows the superposition of two PAM-$4$ signals with adjacent ideal points separated by a distance \(d\). In contrast, subfigure (b) displays the superposition of two skewed QAM-$4$ signals, where the effective Euclidean distances between signal points are increased (e.g., \(d\), \(\sqrt{\tfrac{5}{2}}d\), and \(\sqrt{3}d\)). The skewed QAM-$4$ arrangement achieves an overall larger inter-point distance, enhancing the noise margin relative to the PAM-$4$ configuration.}
    \label{fig:distance}
\end{figure*}
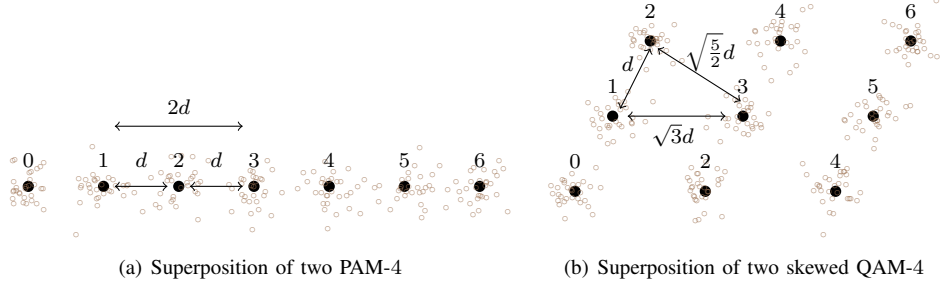

\section{Preliminaries}\label{sec:channelComp}

\subsection{Encoding and Decoding in ChannelComp}

Each digital value ${x}_k$ is selected from finite $q$ possible outputs of $\mathscr{E}_k(\cdot)$. Thus, the received signal $r$  has finite constellation points without channel noise, i.e., $\sum_kx_k$. Indeed, the summation in~\eqref{eq:channelfree} induces a specific constellation diagram of $\sum_kx_k$ that depends on the number of nodes $K$ and on which modulations have been used for each ${x}_k$. Therefore, based on the resultant constellation diagram of signal $r$, one can design a proper decoder (Tabular mapping) $\mathcal{T}({\sum_kx_k})$ to map the constellation points to the output value of the function $f$~\cite{saeed2023ChannelComp}, i.e., we can write this as $\mathcal{T}(\sum_kx_k)=f$.  To illustrate, consider a network of two nodes $K=2$ for computing the max function $f(s_1,s_2)= \max\{s_1, s_2\}$, where the input values are binary, i.e., $s_1,s_2\in\{0,1\}$. In this case, nodes can use simple BPSK modulation to send their data, i.e., 
 \begin{align}
 \label{eq:BPSK}
 {x}_k = \mathscr{E}_k({s}_k) := A_c(2s_k-1), \quad s_k\in \{0,1\},
 \end{align}
 where $A_c$ is the amplitude of the carrier signal. Then, the decoder $\mathscr{D}$, including the ML estimator and tabular map,  is a simple map that assigns values of the constellation diagram of $r$  to the  corresponding output of $f$, i.e.,  
\begin{align}
\label{eq:Example1}
	\mathscr{D}(x) = \begin{cases}
	    1, \quad x \geq -\frac{A_c}{2}, \\
            0, \quad \text{o.w.}
	\end{cases}
\end{align}
This example is illustrated in Figure~\ref{fig:BPSKExample} using blue markers. When the constellation points are distinct, they can be efficiently mapped to the range of the function $f$ by designing an appropriate decoder $\mathscr{D}$. However, Figure~\ref{fig:ConsBPSK} demonstrates a case with $K=2$ nodes computing the max function $f(s_1,s_2) = \max\{s_1,s_2\}$, where $s_1,s_2 \in \{0,1,2,3\}$. In this scenario, the constellation points generated by the nodes cannot be uniquely mapped to the product function. The constellation diagram induced by $\sum_{k=1}^K\mathscr{E}_k({s}_k)$ must span all points in the range of the function ${f}({s}_1,\ldots,s_K)$~\cite{saeed2023ChannelComp}.

The primary challenge lies in designing encoders $\mathcal{E}(\cdot)$ such that the tabular mapper $\mathcal{T}$ can uniquely map each constellation point to a corresponding output value of $f$. Overlapping constellation points, as shown in Figure~\ref{fig:ConsBPSK}, can complicate this process. To address this issue, we devise an optimization framework in the following subsection to avoid overlaps in the constellation diagram.

 \subsection{Resolving Constellation Points Overlaps}

 Let us represent the encoder $\mathcal{E}(\cdot)$ as piecewise map for a given input $s\in \mathcal{X}$, i.e., 
\begin{align}
    \mathcal{E}_k(s) = \sum_{{\ell}=1}^{q}x^{(\ell)}\delta_{a_{\ell}}[s], \quad \forall~a_{\ell}\in \mathcal{X}, \ell \in [q],
\end{align}
where $\delta_{a}[s=a] = 1$ is discrete delta function which one at $s=a_{\ell}$ and zero elsewhere, and  $x^{(\ell)}\in \mathbb{C}$ is complex coefficient of encoder $\mathcal{E}$, which also equals to the constellation point ${\ell}$ of node $k$. In this regard, $\mathcal{E}(\cdot)$ acts as a simple lookup table, where it maps input $s_k$ to the constellation point $x^{(\ell)}$.  Thus, to compute the desired function $f$, we need to obtain all the coefficients $x^{(\ell)}$s to avoid the overlaps among the output values of the functions over the MAC. In fact, let $\mathcal{Y}_f$ and $\mathcal{X}_f$ be the range and domain sets of the function $f$, where $\mathcal{X}_f:=\mathcal{X}_1\times\cdots\mathcal{X}_K$. Also,  let $\mathfrak{I} \subset \mathcal{X}_f$ be a certain combination of the input values  $s_1,s_2\ldots,s_K$ which correspond to the output value $f^{(i)}$.  Then, if $f^{(i)}$ is different from $f^{(j)}$,  the corresponding constellation point ${r}_i:=\sum_{{\ell}\in \mathfrak{I}} x^{(\ell)}$ must not be the same as ${r}_j$ for $i\neq j$. Hence, to verify if  given modulation signals are suitable for computing the desired function $f$, we pose the following problem:
\begin{align}
\label{eq:feasibility}
 \mathcal{P}_1 = {\rm find}  ~~ \bm{x}, ~{\rm s.t.}~ f^{(i)}\neq f^{(i)} \Rightarrow {r}_i \neq {r}_j,~\|\bm{x}\|_2^2 \leq 1,
\end{align}
for all $(i,j) \in \mathcal{Y}_f$, where every element $(i,j)$ is selected as ${f}^{(i)}\neq {f}^{(j)}$, and $\bm{x} \in \mathbb{C}^{q\times 1}$  is the complex modulation vector involving all the constellation points, i.e., 
\begin{align}
    \label{eq:modulation_vector}
    \bm{x} := [x^{(1)},\ldots, x^{(q)}],
\end{align}
in which $x^{(\ell)}$ means $\ell$-th possible constellation points  for $\ell \in [q]$. To avoid such non-smooth constraints in \eqref{eq:feasibility} while avoiding the overlaps, we can solve the following alternative problem to satisfy a Lipschitz smooth condition, i.e., 
\begin{align}
\nonumber
 \mathcal{P}_{2} = &~~~\underset{\bm{x}, \lambda}{\rm max}  ~~ \lambda,  ~~\\ \label{eq:feasibility-conv}
 &{\rm s.t.}~~
\mathcal{D}_{\mathbb{C}}(r_i,r_j) \geq \lambda |f^{(i)}-f^{(j)}|^2, ~\|\bm{x}\|_2^2 \leq 1,
\end{align}
for all $(i,j)\in \mathcal{Y}_f$,  where $\lambda > 0$ is a scaling factor representing the proportionality between the distances. Also, $\mathcal{D}_{\mathbb{C}}$ represents a distance function over complex values. Indeed,  the solution to Problem $\mathcal{P}_1$ equals the solution to Problem $\mathcal{P}_2$~\cite{saeed2023ChannelComp}. Therefore, the problem of finding the encoder to avoid destructive overlaps reduces to optimizing the complex vector $\bm{x}$ that satisfies the constraints in Problem $\mathcal{P}_2$. Another interesting property of the constraints in \eqref{eq:feasibility-conv} is that they are scrutinizing for an encoding that mimics the Lipschitz smoothness of the desired function over the complex plane $\mathbb{C}$. 

However, such a replacement in \eqref{eq:feasibility-conv} imposes a challenge in finding a proper choice for the distance, which we discuss in the sequel.   

\subsection{Distance Choice}

The main question is to find a proper choice of distance function $\mathcal{D}_{\mathbb{C}}$ to obtain the minimum computation error. One might contend that Problem \(\mathcal{P}_2\) could instead be posed with an objective that directly minimizes the average computation error, for example,
\[
\min_{\bm{x}} \; \mathbb{E}\bigl[\,|\,\mathscr{D}(r) - f(s_1,\ldots,s_K)\,|^2\bigr].
\]
However, optimizing solely for average error may permit certain pairs \((i,j)\) with large function differences to collapse in the same constellation points, thereby creating overlaps that persist even in the areas, where the noise variance is low, and cause an irreducible error floor. In contrast, by enforcing the constraint in \eqref{eq:feasibility-conv},  we guarantee a strict separation of all distinct function outputs, preventing destructive collisions. Moreover, selecting \(\mathcal{D}_{\mathbb{C}}\) to reflect the true geometry of the modulation space allows us to balance overlap avoidance with minimization of computation error, yielding robust performance across all different levels of noise. Specifically,  in \cite{saeed2023ChannelComp}, we proposed the following distance functions:
\begin{align}
\label{eq:distnace_euclidean}
    \mathcal{D}_{\mathbb{C}}(r_i,r_j) & = |r_i-r_j|^2,~~\forall r_i,r_j \in \mathbb{C}.
\end{align}
With this choice, we force the constellation points far apart from each other, proportional to the corresponding distance of function values.  Essentially, the spacing between the constellation points mirrors the relative magnitude of the function values. For instance, considering the desired function $f$ as the sum function, the optimization problem in   \eqref{eq:feasibility-conv} would give us pulse amplitude modulation (PAM) as an optimal constellation points diagram.  In particular, $f^{(i)}$ for the sum function can be written as
\begin{align}
    f^{(i)} = \sum_{k=1}^Ks_k^{(i)} = \bm{q}\cdot\bm{a}_i, \quad \bm{q}=[0,1,\ldots,q-1]^{\mathsf{T}}\in \mathbb{Z}^{q},
\end{align}
where $\bm{a}_i\in \mathbb{R}^{q}$ is a binary vector that selects possible cases $i$ of nodes to
send their data to compute $f^{(i)}$. Based on the modulation vector $\bm{x}$ defined in \eqref{eq:modulation_vector}, the constellation $r_i$ is given by $r_i = \bm{x}\cdot\bm{a}_i$ for all $i$. Then, to satisfy the condition in \eqref{eq:feasibility-conv}, one can set $\bm{x}=\sqrt{\lambda} \bm{q}$ which leads to the following 
\begin{align}
    \nonumber
    \mathcal{D}_{\mathbb{C}}(r_i,r_j)  = \lambda \bm{q}^{\mathsf{T}}(\bm{a}_i-\bm{a}_j)(\bm{a}_i-\bm{a}_j)^{\mathsf{T}}\bm{q} = \lambda |f^{(i)}-f^{(j)}|^2,
\end{align}
for all $(i,j)\in \mathcal{Y}_f$.  By adjusting $\lambda$, we can also satisfy the power constraint as well. This essentially suggests PAM as a solution to \eqref{eq:feasibility-conv}. However, it is not difficult to see that the QAM modulation can perform much better than PAM (at least for the scenarios where the noise variance is low), as reflected by the simulation results in \cite{razavikia2023SumCode}. In Figure~\ref{fig:distance}, we illustrate the superposition of two constellation diagrams for PAM-$4$ and QAM-$4$ under the sum function. By preserving a more power-efficient constellation arrangement, QAM increases the distance among adjacent points compared with PAM, thereby reducing computational errors.

This discrepancy raises the question of why ChannelComp suggests such an inefficient solution for computing the sum. Here, we answer this question by considering the channel noise distributions. Indeed, we argue that the problem comes from the chosen Euclidean distance in \eqref{eq:distnace_euclidean}, which ignores the distribution of the channel noise, e.g., Normal distribution. While one might attribute the inefficiency to the inherent NP-hardness of the optimization problem in \eqref{eq:feasibility-conv}, we emphasize that this is not the case for the sum function, for which we have demonstrated optimality of PAM modulation in this work. Hence, we must replace Euclidean distance with a more appropriate noise-aware distance to improve the performance.

\section{Constellation  Design Criteria }\label{sec:DesingCriteria}

In this section, we show how minimizing the computation error over a noisy channel impacts the distance choice and leads to different distance choices when designing the modulation. Moreover, we extend the scheme for different noise distributions over the MAC.

\subsection{Computation by Minimizing Maximum Error}\label{sec:minimizing_maximum_error}

The end goal of the modulation design for the OAC is reliable communication and computation over the noisy channel. Hence, we need to design the constellation diagram of the modulation to minimize the computation error as much as possible. Moreover, as we argued before, the channel noise distribution needs to be considered to quantify the computation error. Consider the MSE of the function $f$ over the channel, i.e., 
\begin{align*}
    \text{MSE}(f) := \mathbb{E}\{|f-\hat{f}|^2\} =\sum_{i}^M \pi_{i}\sum_{i,j}^M \Pr (r_i\rightarrow r_j)|f_i-f_j|^2,
\end{align*}
where $M:= |\mathcal{Y}_f|$ is the cardinality of the range of function $f$, and $\Pr(r_i\rightarrow r_j)$ denotes the probability of detecting $r_j$ instead of $r_i$,  and $\pi_i$ represents the prior probability of transmitting a signal $r_i$ to compute $f_i$.   Under the uniform distribution assumption on the prior probability, it is not difficult to see that the average computation error can be lower and upper bounded by its maximum computation error, i.e., 
\begin{align}
   \label{eq:lower_upper_error}
   \frac{1}{M} \Xi_{\max}  \leq  \text{MSE}(f)  \leq M \Xi_{\max},
\end{align}
where $\Xi_{\max}=\max_{i,j} \Pr(r_i\rightarrow r_j)|f_i-f_j|^2$. Therefore, for a given size of the constellation diagram $M$, the computation error vanishes if and only if the $\Xi_{\max}$ vanishes, i.e., $\lim_{\Xi_{\max}\rightarrow 0}\text{MSE}(f)  =0$.  Hence, our approach minimizes the maximum computation error from detection at the CP to achieve a more reliable computation.  Although one could instead design the constellation by directly minimizing the MSE, such a criterion offers no explicit guarantee against symbol overlap. By contrast, our maximum-error minimization inherently enforces point separation. Moreover, as shown at the end of this section,  the proposed method yields a solution that asymptotically converges to the classical MSE-optimal design under low-variance noise, when the channel is subject to sub-exponential noise distributions.  Particularly, for the probability of the error of computing  $f^{(j)}$ instead of $f^{(i)}$, we can define the computation error between them as  $\Xi_{i,j}:= \Pr(r_i\rightarrow r_j) |f^{(i)}-f^{(j)}|$. Accordingly, to design the constellation diagram,  we pose the following optimization problem 
\begin{align}
     \label{eq:noisy-convex-upper}
    \hat{\bm{x}}= \underset{\bm{x}}{\rm argmin} \max_{i,j}  ~~~ \Xi_{i,j}, ~~~{\rm s.t.}~~~\|\bm{x}\|_{2}^2 \leq  1,
\end{align}
 for all $(i,j)\in [M]\times [M]$. The key challenge in solving~\eqref{eq:noisy-convex-upper} lies in the nonlinearity of $\Pr(r_i\rightarrow r_j)$, which depends on both the channel noise statistics and the modulation geometry. 
  In subsequent subsections, we establish connections between~\eqref{eq:noisy-convex-upper} and the generalized constellation design Problem~$\mathcal{P}_2$, showing how channel-aware distance metrics naturally emerge from the noise distribution. These insights lead to optimized, noise-tailored constellations that enhance computational reliability over noisy MACs.

 \subsection{AWGN Channel}

To solve the min-max optimization problem in~\eqref{eq:noisy-convex-upper}, one must evaluate the probability of symbol misdetection, i.e., $\Pr(r_i\rightarrow r_j)$
 which is generally intractable in closed form.  This difficulty arises because the geometry of the constellation points is unknown before determining the constellation diagram. Consequently, the corresponding decision boundaries are also unavailable, preventing accurate computation of the misdetection probabilities. In the presence of AWGN, a common technique is to upper bound this probability using the union bound: 
 \begin{align} 
 \label{eq:upper_q_normal}
 \Pr(r_i \rightarrow r_j) \leq Q\bigg( \frac{|r_i - r_j|}{\sqrt{2}\sigma}\bigg), 
 \end{align} 
 where $Q(x)$ denotes the Gaussian Q-function defined by $Q(x) = \int_{x}^{\infty}\exp{(-t^2/2)}dt/\sqrt{2\pi}$~\cite{brehler2002asymptotic}. We use the Q function as a surrogate function to replace a simple upper bound with the generally intractable cost function $\Pr(r_i\rightarrow r_j)$. The rationale is that as the Q-function decreases, the probability of error $\Pr(r_i\rightarrow r_j)$ shall also diminish. However, trying to solve optimization problems involving the Q function is still tricky; one well-known way to upper bound the Gaussian Q-function is to use the Chernoff upper bound \cite{Tarokh1998Space}, i.e., $Q(x)\leq {{\rm e}^{-x^2/2}}$ accordingly\footnote{ More accurate approximations than the Chernoff bound are also proposed in \cite{chiani2003new} as the Chiani bound. However, in our optimization problem, the Chiani bound can make the situation more complicated.}. Hence, using the Chernoff bound, the objective function is upper bounded as follows.
\begin{align}
\nonumber
\underset{\bm{x}}{\rm min} \max_{i,j} 
 ~ |f^{(i)}-f^{(j)}|^2~{\rm e}^{-\frac{|r_i-r_j|^2}{4\sigma}}, ~~
{\rm s.t.}~~\|\bm{x}\|_{2}^2 \leq 1, \label{eq:noisy-convex-exp}
\end{align}
Next, because all the terms are strictly positive, by taking the logarithm of the objective function, we could reach
\begin{equation}
    \label{eq:noisy-convex-upper-1}
     \underset{\bm{x}}{\rm min} \max_{i,j}~8\sigma \mu_{i,j} - |r_i-r_j|^2, ~~
    {\rm s.t.}~~
   \|\bm{x}\|_{2}^2 \leq 1,
\end{equation}
where $ \mu_{i,j} := \ln\big(|f^{(i)}-f^{(j)}|\big)$ for $(i,j)\in [M]\times [M]$.  Finally, we can turn the min-max optimization in \eqref{eq:noisy-convex-upper-1} into the following quadratic optimization problem in terms of the modulation vector $\bm{x}$, i.e.,  
\begin{align}
    \label{eq:noisy-convex-upper-2}
    \underset{\bm{x}}{\rm min} ~
 t' ~{\rm s.t.}  ~  t' \geq 8\sigma \mu_{i,j} -|r_i-r_j|^2,~\forall (i,j)  ~ \|\bm{x}\|_{2}^2 \leq  1,
\end{align}
or equivalently, 
\begin{equation}
    \label{eq:noisy-convex-upper-3}
    \underset{\bm{x}}{\rm max}   ~
 t  ~{\rm s.t.}   ~  |r_i-r_j|^2 \geq 8\sigma \mu_{i,j} +t, ~\forall (i,j)~~  \|\bm{x}\|_{2}^2 \leq  1.
\end{equation}
The optimization problem in \eqref{eq:noisy-convex-upper-3} is quadratically constrained in terms of modulation vector $\bm{x}$ and can be solved by relaxation techniques. 

\begin{rem}
  Optimization problem~\eqref{eq:noisy-convex-upper-3} falls under the quadratically constrained quadratic programming class, known as NP-hard~\cite{Sidir2006Physical}. Nevertheless, approximate solutions can be effectively obtained using the lifting trick~\cite{saeed2023ChannelComp},  which reformulates the problem into a higher-dimensional space amenable to convex relaxations.  Also, the corresponding optimality gap associated with this relaxation is characterized in \cite{yan2025remac}. 
\end{rem}

To connect the constraints in \eqref{eq:noisy-convex-upper-3} to the Problem~$\mathcal{P}_2$, we change the variable  $\lambda:= \ln{(t)}/4\sigma$ and rewrite the optimization in \eqref{eq:noisy-convex-upper-3} as follows
\begin{equation}
    \label{eq:awgn_optmization}
    \underset{\bm{x}}{\rm max}   ~
 \lambda  ~{\rm s.t.}   ~  {\rm e}^{|r_i-r_j|^2/4\sigma} \geq  \lambda |f^{(i)}-f^{(j)}|,~~   \|\bm{x}\|_{2}^2 \leq  1,
\end{equation}
for all $(i,j)\in \mathcal{Y}_f$. by comparing the minimization in \eqref{eq:awgn_optmization} and constrains of Problem~$\mathcal{P}_2$ in \eqref{eq:feasibility-conv},  we can see that it suggests that $\mathcal{D}_{\mathbb{C}}$ to be the exponential functions, i.e.,
\begin{align}
   \label{eq:distnaace_exp}
   \mathcal{D}_{\mathbb{C}}(r_i,r_j) = \exp{(|r_i-r_j|^2/4\sigma)}, \quad, \forall r_i,r_j \in \mathbb{C}. 
\end{align}
The proposed distance in \eqref{eq:distnaace_exp} captures the noise distribution characteristics for a computation over a noisy MAC.

\begin{rem}
    The noise variance \(\sigma\) significantly influences the constellation design in \eqref{eq:noisy-convex-upper-3}. In the low-noise regime (\(\sigma \to 0\)), the term \(4\sigma \mu_{i,j}\) becomes negligible, and the optimization prioritizes maximizing the minimum distance \(|r_i - r_j|^2\), resulting in equidistant constellation points akin to traditional communication systems focused on bit error rates. Conversely, in the high-noise regime (\(\sigma \gg 1\)), the term \(8\sigma \mu_{i,j} = 8\sigma \ln(|f^{(i)} - f^{(j)}|)\) dominates, aligning constellation points with the logarithmic differences in function values, emphasizing computational accuracy.
\end{rem}

\begin{rem}
An immediate observation from the formulation in \eqref{eq:noisy-convex-upper-3} is that, although the exact optimal solution remains unknown, one can anticipate that QAM-based constellations may offer more favorable performance than PAM in the low-noise regime (\(\sigma \to 0\)). QAM more effectively maximizes the minimum pairwise Euclidean distance \(|r_i - r_j|^2\) under the power constraint, thereby reducing symbol confusion probability. Consequently, incorporating the noise-aware objective in \eqref{eq:noisy-convex-upper-3} implicitly favors QAM-like geometries that yield enhanced robustness in additive white Gaussian noise.
\end{rem}

\subsection{Extension to Generalized Normal Distribution}\label{sec:Generlzied_normal}

So far, we have assumed an AWGN channel with Gaussian noise, yielding the exponential distance metric \(\mathcal{D}_{\mathbb{C}}(x_1, x_2) = \exp\left(|x_1 - x_2|^2 / 4\sigma\right)\) in \eqref{eq:noisy-convex-upper-3}. Here, we extend the noise distribution to a generalized Normal distribution (GND), characterized by the density \( p_Z(z) = \frac{\beta}{2\alpha \Gamma(1/\beta)} \exp\left(-\left|\frac{z}{\alpha}\right|^\beta\right) \), where \(\beta > 0\) is the shape parameter, \(\alpha > 0\) is the scale, and \(\Gamma\) is the gamma function. This distribution encompasses Gaussian noise (\(\beta = 2\)) and Laplace noise (\(\beta = 1\)) as special cases, enabling the modeling of diverse noise profiles.

We reconsider the error probability \(\Pr(r_i \rightarrow r_j)\) to extend our framework. This probability depends on \(|r_i - r_j|\) and \(\beta\) for GND noise. The right tail probability for $a >0$ is given by
\begin{align}
   \label{eq:error_prob_gamma}
   \Pr(r_i \rightarrow r_j)\leq \frac{\beta}{2\Gamma(1/\beta)}\int_{|r_i - r_j|/\alpha}^{\infty}{\rm e}^{-u^{\beta}}du, 
\end{align}
where the substitution \(u=z/\alpha\) has been applied. A standard inequality for such integrals is
\begin{align}
     \nonumber
    \int_{z}^{\infty}{\rm e}^{-u^{\beta}}\,du \le \frac{1}{\beta\,z^{\beta-1}}{\rm e}^{-z^{\beta}}, \quad z>0.
\end{align}
Setting \(z = |r_i - r_j|/\alpha\), the tail probability can be bounded by
\begin{align*}
    \Pr(r_i \rightarrow r_j) \le \frac{1}{2\Gamma(1/\beta)}\frac{\alpha^{\beta-1}}{|r_i - r_j|^{\beta-1}}{\rm e}^{-\left(\frac{|r_i - r_j|}{\alpha}\right)^{\beta}}, ~~ \forall (i,j).
\end{align*}
 Substituting into \eqref{eq:noisy-convex-upper} and following similar steps, the objective becomes:
\begin{align}
    \underset{\bm{x}}{\rm min} \, \max_{i,j} \, \frac{|f^{(i)} - f^{(j)}|^2}{|r_i-r_j|^{\beta-1}} {\rm e}^{-\frac{|r_i - r_j|^\beta}{\alpha^\beta}}, \quad {\rm s.t.} \quad \|\bm{x}\|_2^2 \leq 1,
\end{align}
This suggests a generalized distance metric:
\begin{align}
 \label{eq:exp_distaces_beta}
    \mathcal{D}_{\mathbb{C}}(r_i, r_j) = |r_i-r_j|^{\beta-1} \exp\left(\frac{|r_i - r_j|^\beta}{\alpha^\beta}\right).
\end{align}
This adaptation highlights the noise tail’s impact on constellation design, controlled by \(\beta\).

\begin{figure}[!t]
  \centering
  \begin{tikzpicture}
     \begin{scope}[spy using outlines={rectangle, magnification=2,
   width=1.5cm,height=1.5cm,connect spies}]
    \begin{axis}[
      width=\linewidth,
      height=0.6\linewidth,
           label style={font=\small},
        tick label style={font=\small},
      xlabel={$x$},
      ylabel={$y$},
      domain=-20:20,
      samples=400,
      legend pos=north west,
      legend cell align=left,
      grid=major,
      xmin=-20, xmax=20,
      ymin=0, ymax=22,
    legend cell align={left},
        legend style={nodes={scale=0.7, transform shape}, at={(0.4,0.95)}},
        ymajorgrids=true,
        xmajorgrids=true,
        grid=both,
        grid style={line width=.1pt, draw=gray!15},
        major grid style={line width=.2pt, draw=gray!40},
    ]

      \addplot[thick, chestnut,line width=1.5pt] {1/(rad(atan(1/abs(x))))};\addlegendentry{$y = {\arctan^{-1}\bigl(1/|x|\bigr)}$}
      \addplot[thick, dashed, airforceblue,line width=1.5pt] {abs(x)};
      \addlegendentry{$y = |x|$}
      \path (0, 1.5) coordinate (X);
    \end{axis}
          \spy [black] on (X) in node (zoom) [left] at ([xshift=0.2cm,yshift=1.5cm]X);
    \end{scope}
  \end{tikzpicture}
  \caption{Comparison of the functions $y=|x|$ (blue) and $y=\arctan^{-1}\bigl(1/|x|\bigr)$ (red) on the interval $[-20,20]$.}
  \label{fig:abs_arctan_reciprocal}
\end{figure}
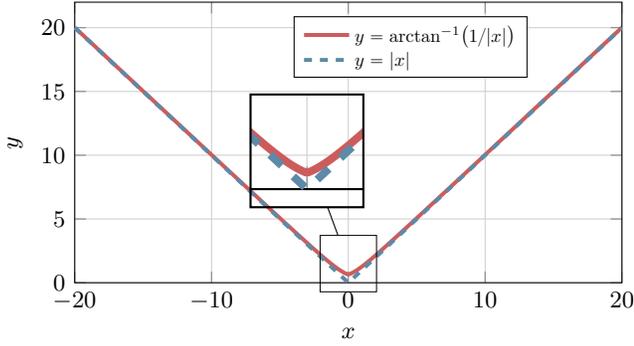

\begin{rem}
The upper bound on the error probability in \eqref{eq:error_prob_gamma} relies on the symmetry of the GND distribution. Specifically, the ML detector under GND noise reduces to the minimum-distance decoder, as the distribution is symmetric about its mean. Consequently, the likelihood is maximized when the received symbol is closest in Euclidean distance to a constellation point, making the decision boundary between any two neighboring symbols their midpoint. This property holds for Gaussian ($\beta=2$) and Laplace ($\beta=1$) cases and justifies using Euclidean distances in bounding error probabilities for symmetric noise distributions.
\end{rem}

The derivation above highlights how noise-tail characteristics—particularly for generalized normal distribution —shape the error bounds and influence modulation design. A notable takeaway is that the resulting surrogate objectives share a standard log-sum-exp structure, asymptotically approximating the MSE as well. To formalize this connection, we next establish that minimizing the MSE becomes approximately equivalent to minimizing the maximum computation error in the low-variance noise regime for sub-exponential noise distributions.


\begin{prop}
\label{prop:subExponential_equiv}
For a communication system in \eqref{eq:channelfree}, if the channel noise $\tilde{z}\sim SE(\nu)$ follows sub-exponential distributions with variance parameter \(\nu\), i.e., 
\begin{align}
    \Pr(|\tilde{z}|\geq t) \leq2 {\rm e}^{-\tfrac{t}{\nu}}, \quad t \geq 0. 
\end{align}
Also, let $\Xi_{i,j}(\bm x)$ be the pairwise computation error as \eqref{eq:noisy-convex-upper}, $\bm{x}$ be the modulation vector,  and the two design objectives
\begin{subequations}
    \begin{align}
\label{eq:opt_mean-prop}
  J_{\rm MSE}(\bm x;\nu)\;&:=\;\sum_{i,j}\Xi_{i,j}(\bm x),
  \\
  J_{\max}(\bm x;\nu)\;&:=\;\max_{i,j}\Xi_{i,j}(\bm x).
\end{align}
\end{subequations}
Then, as $\nu\to 0^{+}$: Let
\begin{align*}
          \bm x_\nu^*
      =\arg\min_{\|\bm x\|\le1}J_{\rm MSE}(\bm x;\nu),
      \quad
      \bm x_0^*
      =\arg\min_{\|\bm x\|\le1}J_{\max}(\bm x;\nu).
\end{align*}
    Then, every accumulation point of $\{\bm x_\nu^*\}_{\nu>0}$ lies in the set of minimizers of $J_{\max}$.  In particular, for sufficiently small $\nu$, any solution of the MSE‐minimization also solves the worst‐case minimization.
\end{prop}

\begin{proof}
    For the proof, see Appendix~\ref{subsec:sub-Exponential}. 
\end{proof}

Proposition~\ref{prop:subExponential_equiv} shows that the design criterion based on minimizing the MSE under sub-exponential channel noise is equivalent to employing a max-min constellation optimization when the channel noise variance is low. In this regime, the rapid tail decay of the sub-exponential distribution leads to dominant errors occurring between nearest-neighbor constellation points (maximum error). Consequently, maximizing the minimum pairwise separation mitigates these dominant errors and reduces the overall computation error.

\subsection{Extension to Heavy-Tailed Noise}\label{sec:Heavy_maxerror}

While the generalized normal distribution accommodates a range of tail behaviors, heavy-tailed noise, prevalent in certain wireless channels, requires further adaptation due to its slow-decaying tails. In practical wireless environments and interference-limited scenarios, the noise exhibits heavy-tailed behavior~\cite{clavier2020experimental}. Consider a Cauchy distribution for the noise, with density
\begin{align}
    p_Z(z)= \frac{1}{\pi \gamma \left(1 + (\frac{z}{\gamma})^2\right)},
\end{align}
where \(\gamma > 0\) is the scale parameter. Unlike Gaussian or GND noise, the Cauchy distribution’s moment-generating function is undefined, rendering the Chernoff bound inapplicable. Instead, we directly analyze the tail probability. For a transmitted signal \(r_i\) and detected signal \(r_j > r_i\), the error probability is given by
\begin{align}
 \nonumber
    \Pr(r_i \rightarrow r_j) &\leq \Pr\left(Z > \frac{|r_j - r_i|}{2}\right), \\
    & =  \frac{1}{2} - \frac{1}{\pi} \arctan\left(\frac{|r_j - r_i|}{2\gamma}\right), \nonumber   \\
    & = \frac{1}{\pi} \arctan\left(\frac{2\gamma}{|r_j - r_i|}\right).\label{eq:Cauchydis}
\end{align}
Since the heavy-tailed distributions do not have a finite variance, MSE is not a suitable choice to minimize. In this regard, we consider the computation error to be $\Xi_{i,j}^{\eta} = \Pr(r_i \rightarrow r_j) |f^{(i)} - f^{(j)}|^{\eta}$ for $\eta <1$. Hence,  by replacing the upper bound on the probability of the error in \eqref{eq:Cauchydis} into the computation error $\Xi_{i,j}^{\eta}$, the optimization becomes:
\begin{align*}
    \underset{\bm{x}}{\rm min} \, \max_{i,j}  \frac{1}{\pi}|f^{(i)} - f^{(j)}|^{\eta}\arctan\left(\frac{2\gamma}{|r_j - r_i|}\right), ~ {\rm s.t.} ~\|\bm{x}\|_2^2 \leq 1.
\end{align*}
Reformulating to maximize the minimum distance adjusted by function differences, we obtain:
\begin{align}
     \nonumber
    \underset{\bm{x}}{\rm max} \, t, ~~~&{\rm s.t.} ~ \arctan^{-1}\left(\frac{2\gamma}{|r_j - r_i|}\right)\geq t |f^{(i)} - f^{(j)}|^{\eta}, \\&~~~~~~~ \|\bm{x}\|_2^2 \leq 1, \label{eq:arctan_constaint}
\end{align}
for all $(i,j)$. Because the inverse‐arctangent constraint is highly nonlinear, which makes solving the optimization in \eqref{eq:arctan_constaint} challenging.  To obtain a tractable surrogate, we use the absolute function as a lower bound for the inverse‐arctangent, i.e., 
\begin{align}
    \arctan^{-1}(1/|x|) \geq |x|, 
\end{align}
which becomes increasingly tight as the input argument grows $x$, i.e., $ x \rightarrow \infty,~ \arctan^{-1}(1/|x|) \approx |x|$. Figure~\ref{fig:abs_arctan_reciprocal} depicts the absolute function and inverse-arctangent. Therefore, by enforcing the simpler linear lower bound on the constant, we obtain the following conservative but tractable optimization problem.
\begin{align}
 \nonumber
    \underset{\bm{x}}{\rm max} \, t, ~{\rm s.t.} ~ \frac{|r_j - r_i|}{2\gamma}\geq t |f^{(i)} - f^{(j)}|^{\eta},\forall (i,j)~~ \|\bm{x}\|_2^2 \leq 1,
\end{align}
yielding a distance metric:
\begin{align}
    \mathcal{D}_{\mathbb{C}}(r_i, r_j) = c_{\gamma,\eta}{|r_i - r_j|^{2/\eta}}, 
\end{align}
where $c_{\gamma,\eta}:={(2\gamma)}^{-2/\eta}$ is a positive constant. For $\eta \rightarrow 1$,  this distance is almost the proposed distance in ChannelComp~\cite{saeed2023ChannelComp}. 
For broader heavy-tailed noise, such as stable distributions with tail decay \(\Pr(Z > a) \sim C_{\alpha}a^{-\alpha}\) (\(\alpha < 2\))\footnote{The tail can be obtained by using Markov’s inequality.  For large enough $x$, one can show $C_{\alpha} \approx 2\Gamma(\alpha) \sin{(\pi\alpha/2)}/\pi $. }, the error probability scales as \(|r_i - r_j|^{\alpha}\), generalizing the metric to
\begin{align}
    \mathcal{D}_{\mathbb{C}}(r_i, r_j) = |r_i - r_j|^{\alpha/\eta}.
\end{align}
This inverse power-law metric underscores the need for significantly larger constellation separations as tail heaviness increases, contrasting with the exponential metrics of lighter-tailed noise.

Up to this point, we have revisited the distance criteria and studied the impact of the channel noise distribution on the choice of distance under the assumption of perfect fading compensation, neglecting any fading phenomena.  In Section~\ref{sec:FadStochastic}, we relax this assumption by introducing a stochastic fading model and demonstrate how random channel variations reshape both the optimization constraints and the resulting distance metric.

\begin{figure*}
\centering
\subfigure[$|r_1-r_2|^2$-$\sum$]{\label{fig:const(a)}
    \begin{tikzpicture}
    \begin{axis}[
        xlabel={$\Re(\bm{x})$},
        ylabel={$\Im(\bm{x})$},
        label style={font=\tiny},
        width=0.25\linewidth,
        height=0.25\linewidth,
        xmin=-0.33, xmax=0.33,
        ymin=-0.33, ymax=0.33,
        ticklabel style={font=\tiny},
        grid=both,
        grid style={line width=.1pt, draw=gray!10},
        major grid style={line width=.2pt,draw=gray!30},
    ]
    \addplot[
        color=charcoal,
        mark=triangle,
        mark options={rotate=180},
        line width=1pt,
        mark size=2pt,
    ]
    table[col sep=space, x=ChannelComp_x, y=ChannelComp_y] {Data/constellation_sum_5bits.dat};
    \end{axis}
    \end{tikzpicture}
}\subfigure[$|r_1-r_2|$-$\sum$]{
    \begin{tikzpicture}
    \begin{axis}[
        xlabel={$\Re(\bm{x})$},
        ylabel={$\Im(\bm{x})$},
        label style={font=\tiny},
        width=0.25\linewidth,
        height=0.25\linewidth,
        xmin=-0.22, xmax=0.22,
        ymin=-0.22, ymax=0.22,
        ticklabel style={font=\tiny},
        grid=both,
        grid style={line width=.1pt, draw=gray!10},
        major grid style={line width=.2pt,draw=gray!30},
    ]
    \addplot[
        color=darkcerulean,
        mark=square,
        line width=1pt,
        mark size=2pt,
    ]
    table[col sep=space, x=heavy_x, y=heavy_y] {Data/constellation_sum_5bits.dat};
    \end{axis}
    \end{tikzpicture}
}\subfigure[$\exp{({|r_1-r_2|})}$-$\sum$]{
    \begin{tikzpicture}
    \begin{axis}[
        xlabel={$\Re(\bm{x})$},
        ylabel={$\Im(\bm{x})$},
        label style={font=\tiny},
        width=0.25\linewidth,
        height=0.25\linewidth,
        xmin=-0.2, xmax=0.2,
        ymin=-0.2, ymax=0.2,
        ticklabel style={font=\tiny},
        grid=both,
        grid style={line width=.1pt, draw=gray!10},
        major grid style={line width=.2pt,draw=gray!30},
    ]
    \addplot[
        color=chamoisee,
        mark=diamond,
        line width=1pt,
        mark size=2pt,
    ]
    table[col sep=space, x=Laplace_x, y=Laplace_y] {Data/constellation_sum_5bits.dat};
    \end{axis}
    \end{tikzpicture}
}\subfigure[$\exp{({|r_1-r_2|^2})}$-$\sum$]{\label{fig:const(d)}
    \begin{tikzpicture}
    \begin{axis}[
        xlabel={$\Re(\bm{x})$},
        ylabel={$\Im(\bm{x})$},
        label style={font=\tiny},
        width=0.25\linewidth,
        height=0.25\linewidth,
        xmin=-0.2, xmax=0.2,
        ymin=-0.2, ymax=0.2,
        ticklabel style={font=\tiny},
        grid=both,
        grid style={line width=.1pt, draw=gray!10},
        major grid style={line width=.2pt,draw=gray!30},
    ]
    \addplot[
        color=eggplant,
        mark=o,
        line width=1pt,
        mark size=2pt,
    ]
    table[col sep=space, x=AWGN_x, y=AWGN_y] {Data/constellation_sum_5bits.dat};
    \end{axis}
    \end{tikzpicture}
}

\subfigure[$|r_1-r_2|^2$-$\prod$]{\label{fig:const(e)}
    \begin{tikzpicture}
    \begin{axis}[
        xlabel={$\Re(\bm{x})$},
        ylabel={$\Im(\bm{x})$},
        label style={font=\tiny},
        width=0.25\linewidth,
        height=0.25\linewidth,
       xmin=-0.28, xmax=0.28,
        ymin=-0.28, ymax=0.28,
        ticklabel style={font=\tiny},
        grid=both,
        grid style={line width=.1pt, draw=gray!10},
        major grid style={line width=.2pt,draw=gray!30},
    ]
    \addplot[
        color=charcoal,
        mark=triangle,
        mark options={rotate=180},
        line width=1pt,
        mark size=2pt,
    ]
    table[col sep=space, x=ChannelComp_x, y=ChannelComp_y] {Data/constellation_prod_5bits.dat};
    \end{axis}
    \end{tikzpicture}
}\subfigure[$|r_1-r_2|$-$\prod$]{
    \begin{tikzpicture}
    \begin{axis}[
        xlabel={$\Re(\bm{x})$},
        ylabel={$\Im(\bm{x})$},
        label style={font=\tiny},
        width=0.25\linewidth,
        height=0.25\linewidth,
        xmin=-0.3, xmax=0.3,
        ymin=-0.3, ymax=0.3,
        ticklabel style={font=\tiny},
        grid=both,
        grid style={line width=.1pt, draw=gray!10},
        major grid style={line width=.2pt,draw=gray!30},
    ]
    \addplot[
        color=darkcerulean,
        mark=square,
        line width=1pt,
        mark size=2pt,
    ]
    table[col sep=space, x=heavy_x, y=heavy_y] {Data/constellation_prod_5bits.dat};
    \end{axis}
    \end{tikzpicture}
}\subfigure[$\exp{({|r_1-r_2|})}$-$\prod$]{
    \begin{tikzpicture}
    \begin{axis}[
        xlabel={$\Re(\bm{x})$},
        ylabel={$\Im(\bm{x})$},
        label style={font=\tiny},
        width=0.25\linewidth,
        height=0.25\linewidth,
        xmin=-0.28, xmax=0.28,
        ymin=-0.28, ymax=0.28,
        ticklabel style={font=\tiny},
        grid=both,
        grid style={line width=.1pt, draw=gray!10},
        major grid style={line width=.2pt,draw=gray!30},
    ]
    \addplot[
        color=chamoisee,
        mark=diamond,
        line width=1pt,
        mark size=2pt,
    ]
    table[col sep=space, x=Laplace_x, y=Laplace_y] {Data/constellation_prod_5bits.dat};
    \end{axis}
    \end{tikzpicture}
}\subfigure[$\exp{({|r_1-r_2|^2})}$-$\prod$]{\label{fig:const(h)}
    \begin{tikzpicture}
    \begin{axis}[
        xlabel={$\Re(\bm{x})$},
        ylabel={$\Im(\bm{x})$},
        label style={font=\tiny},
        width=0.25\linewidth,
        height=0.25\linewidth,
        xmin=-0.25, xmax=0.25,
        ymin=-0.25, ymax=0.25,
        ticklabel style={font=\tiny},
        grid=both,
        grid style={line width=.1pt, draw=gray!10},
        major grid style={line width=.2pt,draw=gray!30},
    ]
    \addplot[
        color=eggplant,
        mark=o,
        line width=1pt,
        mark size=2pt,
    ]
    table[col sep=space, x=AWGN_x, y=AWGN_y] {Data/constellation_prod_5bits.dat};
    \end{axis}
    \end{tikzpicture}
}
\caption{Figure\ref{fig:const(a)}-\ref{fig:const(d)} depict constellation diagrams for $f(\bm{x})=\sum_{k=1}^{K} x_k$ with $K=2$ nodes and $q=32$ (5 bits). Figure\ref{fig:const(e)}-\ref{fig:const(h)} 
 shows the constellation diagrams for $f(\bm{x})=\prod_{k=1}^{K} x_k$ with $K=2$ nodes and $q=32$ (5 bits). Each subfigure corresponds to a different distance function.}
\label{fig:ConstellationDiagram}
\end{figure*}
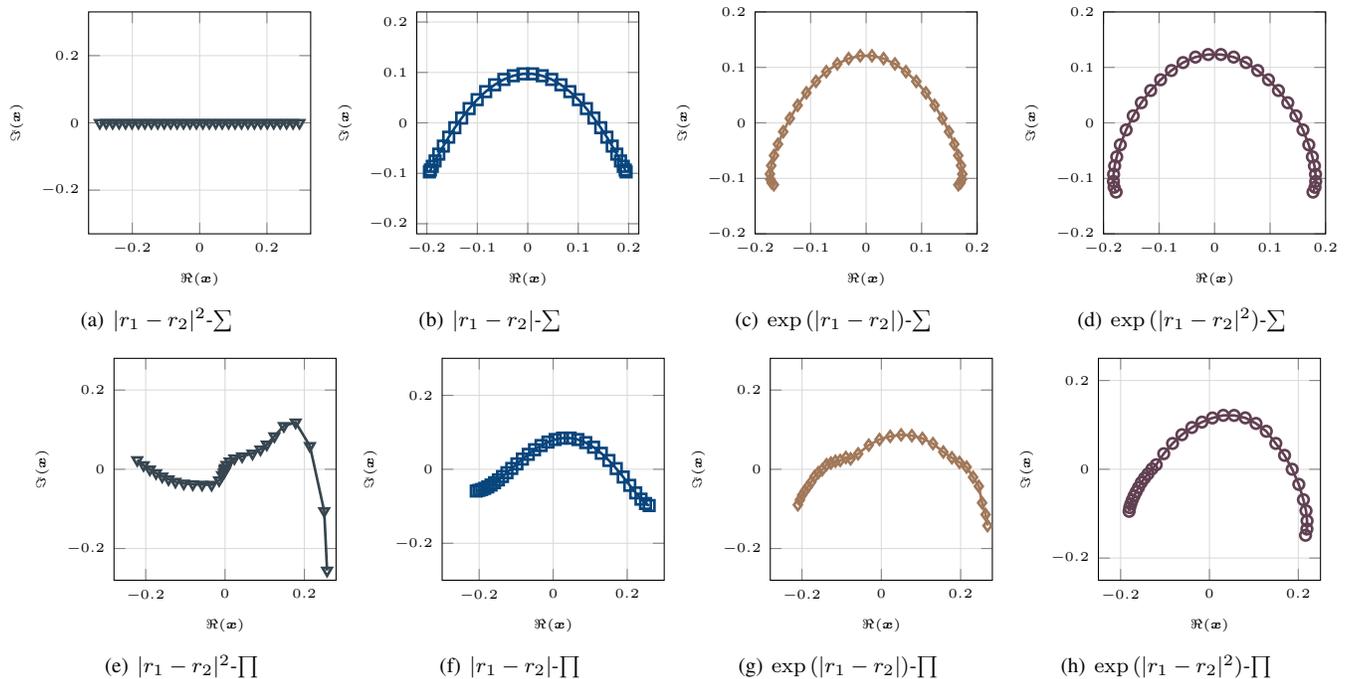

\subsection{Stochastic Fading Channel}\label{sec:FadStochastic}

In many wireless scenarios, channel coefficients $h_k$ can be estimated and used imperfectly during communication. However, in some cases, direct channel estimation is infeasible. Instead, we may characterize the channel statistically by modeling its distribution. A widely adopted model is Rician fading, which captures environments with a dominant line-of-sight component. Without such a component, the model reduces to Rayleigh fading.

When $h_k$ becomes a random variable, the optimization problem associated with the digital OAC becomes stochastic. A common approach is to replace the channel-dependent constraints with their expectations. To be more precise, let $\bm{h} := [h_1,\ldots,h_K]^{\mathsf{T}} \in \mathbb{C}^{K}$ be the random channel vector. Then, the resultant constellation point $r_i$ is expressed as
\begin{align}
    r_i = \bm{h}^{\mathsf{T}}\bm{B}_i \bm{x}, \quad \forall\,i\in[M],
\end{align}
where $\bm{B}_i \in \{0,1\}^{K\times q}$ is a binary selection matrix indicating the combination of the constellation points from the modulation vector $\bm{x}$ which results in $r_i$. The difference between any two constellation points $r_i$ and $r_j$ can be written as
\begin{align}
    |r_i - r_j|^2 = \bm{x}^{\mathsf{H}} \tilde{\bm{B}}_{i,j} \bm{h} \bm{h}^{\mathsf{H}} \tilde{\bm{B}}_{i,j}^{\mathsf{H}} \bm{x},
\end{align}
where $\tilde{\bm{B}}_{i,j} = \bm{B}_i - \bm{B}_j$. Taking the expectation yields
\begin{align}
    \nonumber
\mathbb{E}\Big[|r_i-r_j|^2\Big] & =\mathbb{E}\Big[\bm{x}^{\mathsf{H}}\tilde{\bm{B}}_{i,j}\bm{h}\bm{h}^{\mathsf{H}}\tilde{\bm{B}}_{i,j}^{\mathsf{H}}\bm{x}\Big], \\ \nonumber
& = \bm{x}^{\mathsf{H}}\tilde{\bm{B}}_{i,j}\mathbb{E}[\bm{h}\bm{h}^{\mathsf{H}}]\tilde{\bm{B}}_{i,j}^{\mathsf{H}}\bm{x}, \\ 
& = \bm{x}^{\mathsf{H}}\tilde{\bm{B}}_{i,j}\bm{K}\tilde{\bm{B}}_{i,j}^{\mathsf{H}}\bm{x}, \label{eq:Stochastic}
\end{align}
where $\bm{K} := \mathbb{E}[\bm{h}\bm{h}^{\mathsf{H}}] \in \mathbb{C}^{K\times K}$ is the channel covariance matrix, with entries $\rho_{k,k'} = \mathbb{E}[h_k h_{k'}^*]$. 

Hence, the effect of fading can be incorporated by replacing $\bm{A}_{i,j}$ in Section~\ref{sec:DesingCriteria} with $\tilde{\bm{B}}_{i,j} \bm{K} \tilde{\bm{B}}_{i,j}^{\mathsf{H}}$, leaving the rest of the optimization framework unchanged. This substitution ensures that the resulting modulation scheme is optimized in expectation. Note, however, that the computed function value is not guaranteed to be exact in each realization but is statistically reliable on average.

\begin{figure*}[t]
\centering
\subfigure[$\sum$]{
    \label{fig:MSE_sum}
    \begin{tikzpicture}
    \begin{axis}[
        xlabel={$\sigma$},
        ylabel={MSE (dB)},
        label style={font=\scriptsize},
        tick label style={font=\scriptsize},
        width=0.48\textwidth,
        height=5cm,
        xmin=0.05, xmax=5,
        minor tick num=5,
        ymin=-15, ymax=15,
        legend cell align={left},
        legend style={nodes={scale=0.85, transform shape}, at={(0.98,0.5)}},
        ymajorgrids=true,
        xmajorgrids=true,
        grid=both,
        grid style={line width=.1pt, draw=gray!15},
        major grid style={line width=.2pt, draw=gray!40},
    ]
    \addplot[
        color=charcoal,
        mark=star,
        line width=1pt,
        mark size=2pt,
        smooth,
    ]
    table[x=SNR,y=MSE_sum_ChannelComp]{Data/mse_Normal_sum.dat};
    \addplot[
        color=darkcerulean,
        mark=diamond,
        line width=1pt,
        mark size=2.5pt,
        smooth,
    ]
    table[x=SNR,y=MSE_sum_heavy]{Data/mse_Normal_sum.dat};
    \addplot[
        color=chamoisee,
        mark=square,
        line width=1pt,
        mark size=2pt,
        smooth,
    ]
    table[x=SNR,y=MSE_sum_Laplace]{Data/mse_Normal_sum.dat};
    \addplot[
        color=eggplant,
        mark=o,
        line width=1pt,
        mark size=2pt,
        smooth,
    ]
    table[x=SNR,y=MSE_sum_AWGN]{Data/mse_Normal_sum.dat};
    \legend{ ChannelComp,   Heavy-tail,  Laplace, AWGN};
    \end{axis}
    \end{tikzpicture}
}
\subfigure[$\max$]{
    \label{fig:MSE_prod}
    \begin{tikzpicture}
    \begin{axis}[
        xlabel={$\sigma$},
        ylabel={MSE (dB)},
        label style={font=\scriptsize},
        tick label style={font=\scriptsize},
        width=0.48\textwidth,
        height=5cm,
        xmin=0.08, xmax=5,
        minor tick num=5,
        ymin=-5, ymax=7,
        legend cell align={left},
        legend style={nodes={scale=0.85, transform shape}, at={(0.98,0.5)}},
        ymajorgrids=true,
        xmajorgrids=true,
        grid=both,
        grid style={line width=.1pt, draw=gray!15},
        major grid style={line width=.2pt, draw=gray!40},
    ]
        \addplot[
        color=darkcerulean,
        mark=diamond,
        smooth,
        line width=1pt,
        mark size=2.5pt,
    ]
    table[x=SNR,y=MSE_prod_heavy]{Data/MSEOP_prod.dat};
        \addplot[
        color=charcoal,
        mark=star,
        smooth,
        line width=1pt,
        mark size=2pt,
    ]
    table[x=SNR,y=MSE_prod_ChannelComp]{Data/MSEOP_prod.dat};
    \addplot[
        color=chamoisee,
        mark=square,
        line width=1pt,
        mark size=2pt,
        smooth,
    ]
    table[x=SNR,y=MSE_prod_Laplace]{Data/MSEOP_prod.dat};
    \addplot[
        color=eggplant,
        mark=o,
        line width=1pt,
        smooth,
        mark size=2pt,
    ]
    table[x=SNR,y=MSE_prod_AWGN]{Data/MSEOP_prod.dat};
    \legend{Heavy-tail, ChannelComp,  Laplace,AWGN };
    \end{axis}
    \end{tikzpicture}
}
\caption{MSE performance for distribution‐aware constellation designs over a noisy MAC with Gaussian noise.  Figure~\ref{fig:MSE_sum} and \ref{fig:MSE_prod} present results for the sum function and product functions, respectively.  The figures are depicted for  $10^4$ Monte Carlo trial simulations on a network of $K=12$ nodes with modulation of order $q=16$. }
\label{fig:MSE_AWGN}
\end{figure*}
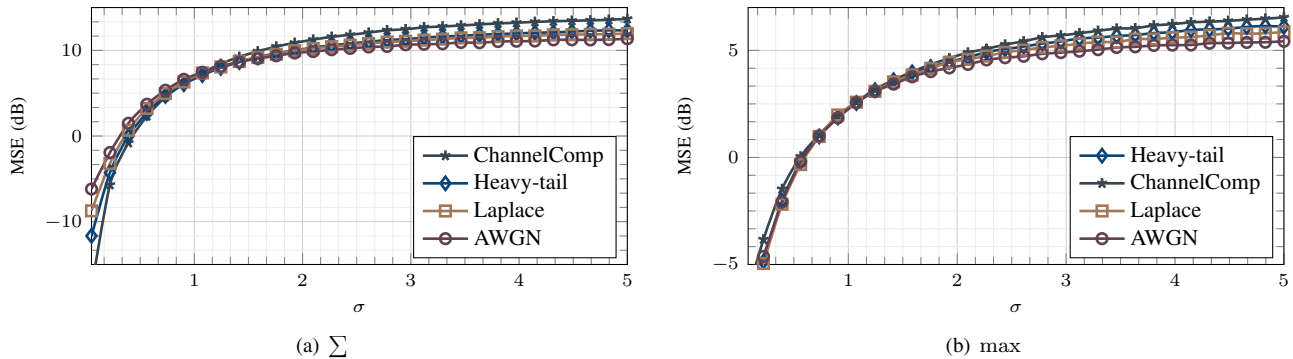

\section{Performance Evaluation}\label{sec:Evaluation}

In this section, we evaluate the performance of the proposed distances in Section~\ref{sec:DesingCriteria} over a noisy MAC with different noise distributions. We also study the impact of the various distances on the modulation design for computing specific functions. Moreover, we empirically compare the computation performance of this designed modulation in terms of the MSE. Throughout this section, to evaluate the performance,  we use the MSE metric, which is defined as 
\begin{align*}
    {\rm MSE}:= \sum_{j=1}^{N_s}\frac{|f^{(i)} - \hat{f}_j^{(i)}|^2}{N_s},
\end{align*}
where $N_s$ denotes the number of Monte Carlo trials, $f^{(i)}$ denotes the value of the desired function we wish to compute, and $\hat{f}_i^{(j)}$ is the $j$-th estimated value of $f^{(i)}$ for $j\in [N_s]$.

\subsection{Distribution Aware Modulation Design}

We provide the modulation design by solving the optimization problem $\mathcal{P}_2$ under various channel noise models discussed in Section~\ref{sec:DesingCriteria}. Specifically, we solve Problem~$\mathcal{P}_2$ for $K=2$ number of nodes and $q=32$ order of the modulation for different distances listed in Table~\ref{tab:distances}.  
Figure~\ref{fig:const(a)}–\ref{fig:const(d)} and \ref{fig:const(e)}–\ref{fig:const(h)} illustrate the resulting constellation diagrams for computing the sum and product functions, respectively. Each subfigure corresponds to a different distance choice. The first row subfigures show the constellation diagrams obtained for computing the function sum $f=\sum_{k}s_k$, whereas the product function $f=\prod_{k}s_k$ is shown in the second row subfigures.

The design is guided by distribution-aware distance metrics listed in Table~\ref{tab:distances}. For instance, under the AWGN model, the proposed method favors a more compact constellation diagram than the other distances for both sum and product functions. On the other hand, ChannelComp's diagram tends to set the constellation points far apart, as it is more suitable for uniform or heavy trial distributions to minimize the computation error. Laplace and AWGN-based distances both provide similar constellation diagrams,  which comes from the fact that both exponentially spaced constellation points as derived in \eqref{eq:distnaace_exp} and \eqref{eq:exp_distaces_beta}, whereas, for ChannelComp and Cauchy noise with $\eta=2$ (since the metric is MSE here), the design adapts to heavier-tailed behavior using polynomial and power-law distances, respectively. These tailored constellations ensure that the pairwise distances between constellation points align with the noise distribution, effectively minimizing computation error.

\begin{table}[t]
\centering
\caption{Distribution-Aware Distance Metrics Used in $\mathcal{P}_2$}
\begin{tabular}{|c|c|}
\hline
\textbf{Noise Distribution} & \textbf{Distance Metric $\mathcal{D}_{\mathbb{C}}(r_1, r_2)$} \\
\hline
AWGN (Gaussian) & $\exp\left( \frac{|r_1 - r_2|^2}{4\sigma} \right)$ \\
\hline
Laplace & $ \exp\left( {|r_1 - r_2|} \right)$ \\
\hline
ChannelComp  & $|r_1 - r_2|^{2}$ \\
\hline
Heavy-Tailed & $|r_1 - r_2|$ \\
\hline
\end{tabular}\label{tab:distances} 
\end{table}

\begin{figure*}[t]
\centering
\subfigure[$\sum_kx_k/K$]{
    \label{fig:MSE_Heavy(a)}
    \begin{tikzpicture}
    \begin{axis}[
        xlabel={$\gamma$},
        ylabel={MAE (dB)},
        label style={font=\scriptsize},
        tick label style={font=\scriptsize},
        width=0.48\textwidth,
        height=5cm,
        xmin=0.05, xmax=1,
        ymax=1.5,
        minor tick num=5,
        legend cell align={left},
        legend style={nodes={scale=0.85, transform shape}, at={(0.98,0.55)}},
        ymajorgrids=true,
        xmajorgrids=true,
        grid=both,
        grid style={line width=.1pt, draw=gray!15},
        major grid style={line width=.2pt, draw=gray!40},
    ]
    \addplot[
        color=eggplant,
        mark=o,
        line width=1pt,
        mark size=2pt,
    ]
    table[x=SNR,y=MSE_sum_AWGN]{Data/MSEOHeavy_Sum.dat};
    \addplot[
        color=chamoisee,
        mark=square,
        line width=1pt,
        mark size=2pt,
    ]
    table[x=SNR,y=MSE_sum_Laplace]{Data/MSEOHeavy_Sum.dat};
    \addplot[
        color=darkcerulean,
        mark=diamond,
        line width=1pt,
        mark size=2.5pt,
    ]
    table[x=SNR,y=MSE_sum_heavy]{Data/MSEOHeavy_Sum.dat};

    \addplot[
        color=charcoal,
        mark=star,
        line width=1pt,
        mark size=2pt,
    ]
    table[x=SNR,y=MSE_sum_ChannelComp]{Data/MSEOHeavy_Sum.dat};
    \legend{ AWGN,  Laplace, Heavy-tail, ChannelComp};
    \end{axis}
    \end{tikzpicture}
}
\subfigure[$(\prod_kx_k)^{1/K}$]{
    \label{fig:MSE_Heavy(b)}
    \begin{tikzpicture}
    \begin{axis}[
        xlabel={$\rm \gamma $},
        ylabel={MAE (dB)},
        label style={font=\scriptsize},
        tick label style={font=\scriptsize},
        width=0.48\textwidth,
        height=5cm,
        xmin=0.1, xmax=1,
        ymin=-3, 
         minor tick num=5,
          y filter/.code={\pgfmathparse{10*log10(\pgfmathresult)}},
        legend cell align={left},
        legend style={nodes={scale=0.85, transform shape}, at={(0.98,0.55)}},
        ymajorgrids=true,
        xmajorgrids=true,
        grid=both,
        grid style={line width=.1pt, draw=gray!15},
        major grid style={line width=.2pt, draw=gray!40},
    ]
    \addplot[
        color=eggplant,
        mark=o,
        line width=1pt,
        mark size=2pt,
    ]
    table[x=SNR,y=MSE_prod_AWGN]{Data/MSEOHeavy_Prod.dat};
    \addplot[
        color=chamoisee,
        mark=square,
        line width=1pt,
        mark size=2pt,
    ]
    table[x=SNR,y=MSE_prod_Laplace]{Data/MSEOHeavy_Prod.dat};
    \addplot[
        color=darkcerulean,
        mark=diamond,
        line width=1pt,
        mark size=2.5pt,
    ]
    table[x=SNR,y=MSE_prod_heavy]{Data/MSEOHeavy_Prod.dat};
    \addplot[
        color=charcoal,
        mark=star,
        line width=1pt,
        mark size=2pt,
    ]
    table[x=SNR,y=MSE_prod_ChannelComp]{Data/MSEOHeavy_Prod.dat};
    \legend{ AWGN, Laplace, Heavy-tail,ChannelComp};
    \end{axis}
    \end{tikzpicture}
}
\caption{MSE performance for distribution-aware constellation designs over a noisy MAC with Cauchy distribution. Figure~\ref{fig:MSE_Heavy(a)} presents results for the Arithmetic mean function, and Figure~\ref{fig:MSE_Heavy(b)} for the Geometric mean function, obtained from $5\times 10^3$ Monte Carlo simulations on a network of $6$ nodes with modulation of order $q=16$. The curves compare designs employing AWGN and Laplace distance metrics—characterized by exponential constellation spacing—with those based on ChannelComp and heavy‐tail metrics with $\eta=2$.}
\label{fig:MSE_Heavy}
\end{figure*}
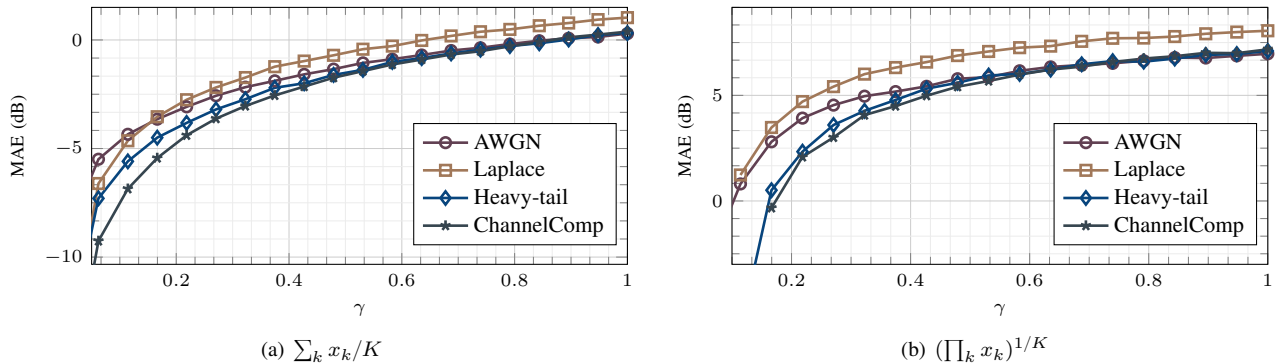

\subsection{Computing Over AWGN}

For the second experiment,  we consider two desired functions: (i) the sum function $f(\bm{x})=\sum_{k}x_k$ and (ii) the product $f(\bm{x})=\max_{k}x_k$, where each function is computed over a network of $K=12$ nodes. Each node is assigned $4$ bits per sample, resulting in $q = 2^4$ possible quantization levels.  The constellation diagram is obtained by solving the SDP relaxation for Problem~$\mathcal{P}_2$ with different distances listed in Table~\ref{tab:distances}. We note that to handle computational complexity for solving the optimization  Problem~$\mathcal{P}_2$, we use the sampling strategy proposed in \cite{razavi2025revisitII} for $p=3$. The computation is performed over the noisy channel as in \eqref{eq:channelfree} where $\tilde{z}\sim \mathcal{CN}(0,\sigma^2)$  for a range of $\sigma\in [0,1]$.  The performance of the obtained constellations is then evaluated via $10^4$  Monte Carlo trials in terms of MSE.

Figures~\ref{fig:MSE_sum} and~\ref{fig:MSE_prod} present the simulation results for the sum and max functions, respectively. For the sum function, in the high noise variance regime ($\sigma > 1$), the exponential distance metric derived under the AWGN assumption demonstrates superior performance compared to other methods, while the Euclidean metric used by ChannelComp yields the poorest results. Conversely, in the low noise variance regime ($\sigma < 1$), the performance ranking is reversed: the Euclidean metric employed by ChannelComp outperforms the others, and the exponential metrics perform suboptimally. This behavior is primarily attributed to the upper bound expression in~\eqref{eq:upper_q_normal}, where the decision boundaries of the superimposed constellation, i.e., $\sum_kx_k$, points for the sum function become nearly vertical. This geometry indicates that only a single access dimension significantly contributes to the decision process. While the exponential metric increases the Euclidean separation between constellation points, it neglects the structure of the actual decision regions. As a result, the effective separation along the true decision boundaries is diminished. Hence, for accurate performance characterization, distance metrics should be evaluated along the real or imaginary axes, aligned with the orientation of the decision regions.  

For the max function computing in Figure~\ref{fig:MSE_prod}, we observe a consistent trend,  where AWGN and Laplace-based show superior performance and less MSE compared to the other distance metrics designs, across different variances of the channel noise.   

Overall, these results confirm that careful constellation design—incorporating an appropriate distance metric through weight functions—can substantially enhance the reliability of function computation over the AWGN channel.

\subsection{Computing Over Heavy Tail Distributions}

For the final experiment, we evaluate the robustness of the proposed design under heavy-tailed noise models, explicitly using the Cauchy distribution with varying scale parameters $\gamma \in [0,1]$.  This setting captures the presence of substantial impulsive interference and models environments where the noise has infinite variance, posing significant challenges for reliable computation.

 In this experiment, the desired functions are Arithmetic function $\sum_kx_k/K$ and Geometric function  $(\prod_kx_k)^{K}$, with $K=6$ nodes and $q=16$ bits per sample, and solve the optimization problem $\mathcal{P}_2$ using the distance metric derived for Cauchy noise in Section~\ref{sec:Heavy_maxerror}. The function computations are carried over noisy MAC, where $z= z_1+z_2{\rm j}\in \mathbb{C},~ z_1,z_2 \sim \text{ Cauchy}(0,\gamma/2)$ for a wide range of $\gamma$.   

The results, presented in Figure~\ref{fig:MSE_Heavy(a)} and Figure~\ref{fig:MSE_Heavy(b)}, demonstrate that the Cauchy-aware constellation design significantly outperforms the other constellation diagrams over a wide range of $\gamma$. In particular, the performance gain becomes more pronounced as the tail of the noise distribution becomes heavier (i.e., larger $\gamma$). Indeed, we observe a similar trend for mean and geometric functions, where ChannelComp and Heavy-Tail outperform the other schemes thanks to the distribution-aware design.

These findings highlight that the proposed design scheme for the constellation diagram, according to the tails of the channel noise distributions, is effective for low-variance noise channel regions. The constellation diagram performs computations based on the channel conditions. Indeed, this emphasizes the value of tailoring the constellation geometry to the statistical structure of the channel noise.

\section{Conclusion}\label{sec:Conclusion}

In this work, we revisited the design of modulation schemes for digital over-the-air computation (OAC) and substantially generalized the ChannelComp framework to achieve robust function evaluation over noisy MACs. We formulated constellation design as a max–min optimization problem that explicitly incorporates the underlying noise distribution through tailored distance metrics. Our analysis reveals that minimizing the maximum computation error yields distinct constellation geometries—each tailored to the specific noise distributions, such as Gaussian, Laplace, or heavy-tailed Cauchy noise—by enforcing appropriate symbol‐to‐function‐value spacings. We further uncover a fundamental connection between the ChannelComp formulation and heavy‐tailed noise behaviors. Extensive simulations demonstrate that the proposed digital scheme consistently outperforms analog and existing digital baselines across low‐ and high‐variance noise regimes, enabling practical, scalable OAC for emerging wireless applications such as edge computing, federated learning, and IoT.

While the proposed framework provided promising results, it relies on several analytical approximations. A rigorous theoretical treatment is necessary to establish more general conclusions, particularly regarding misdetection probability. Our observations in the context of sum computation indicate that solely accounting for the channel noise distribution is insufficient; additional structural properties must be considered. Future work will explore design strategies that incorporate the geometric characteristics of the target function into the constellation optimization process.

In Part II, we tackle the computational burden of modulation optimization by introducing a sampling‐based algorithm and a constraint‐reduction method that exploits symmetry in the target function. This effectively alleviates the complexity inherent in digital encoder design.

\appendix

\subsection{Proof of Proposition~\ref{prop:subExponential_equiv}}\label{subsec:sub-Exponential}

In this section, we prove that minimizing the MSE becomes approximately equivalent to minimizing the maximum error for small noise variances under a \emph{sub-exponential} noise distribution. Specifically, when designing modulation points $\{r_i\}$ for sub-exponential noise, we can employ an MSE-based criterion that asymptotically matches the maximum-error criterion as variance diminishes. More precisely,  assume a uniform prior distribution on symbols, i.e., $\pi_i = 1/M$ for $i \in [M]$. The MSE objective is
\begin{align}
\label{eq:opt_mean}
    \hat{\bm{x}} = \underset{\bm{x}}{\rm argmin}\;\;\sum_{i,j}\,\Xi_{i,j},
    \quad \text{s.t.}\quad 
    \|\bm{x}\|_{2}^2 \;\le\;1,
\end{align}
where $\Xi_{i,j} = \Pr(r_i \to r_j)\,\lvert f_i - f_j\rvert^2$ and $r_i$ is the noiseless received value for symbol $f_i$. 

Similar to Section~\ref{sec:minimizing_maximum_error}, the main challenge lies in the nonlinear term $\Pr(r_i \to r_j)$. For sub-exponential noise, however, we can bound this probability by an exponential function of the absolute distance $\lvert r_i - r_j\rvert$. Specifically, recall that a random variable $Z$ is sub-exponential if there exists $\nu>0$ such that  
\begin{align}
    \label{eq:propSubExponential}
    \Pr\bigl(r_i \to r_j\bigr) 
    \le  
    \Pr\!\Bigl(Z \,\ge\, \tfrac{\lvert r_i - r_j\rvert}{2}\Bigr)
    \le
    2 {\rm e}^{-\,\frac{\lvert r_i - r_j\rvert}{\nu}},
\end{align}
for all $(i,j)$ such that $r_i\neq r_j$. This captures the tail decay rate inherent to sub-exponential noise. Substituting~\eqref{eq:propSubExponential} into \eqref{eq:opt_mean} provides an upper bound for $\Xi_{i,j}$, leading to the following surrogate:
\begin{align}
    \label{eq:opt_mean_surogate}
    \hat{\bm{x}}=
    \underset{\|\bm{x}\|_2^2 \le 1}{\mathrm{argmin}}
    \sum_{i,j}
    {\exp}\Big({-\tfrac{\lvert r_i - r_j\rvert}{\nu}}\Big)\lvert f_i - f_j\rvert^2.
\end{align}
Next, we can write $r_i = \bm{a}_i^\mathsf{T}\bm{x}$, where $\bm{a}_i$ is a binary vector of size $q\times 1$ whose elements are arranged in a way to produce the constellation points $r_i$. Then, we write $\lvert r_i - r_j\rvert = \bigl\lvert \langle\bm{a}_i - \bm{a}_j, \bm{x}\rangle\bigr\rvert$. Also, by recasting $\lvert f_i - f_j\rvert^2$ inside the exponent via taking a logarithm from the objective in \eqref{eq:opt_mean_surogate}, we obtain
\begin{align}
    \label{eq:opt_mean_surogate_2}
    \hat{\bm{x}}
    =
    \underset{\|\bm{x}\|\le 1}{\mathrm{argmin}}
    \sum_{i,j}{\exp}{\Big(
      -\tfrac{\bigl\lvert \langle\bm{a}_i - \bm{a}_j, \bm{x}\rangle\bigr\rvert}{\nu}
      +
      2\ln\lvert f_i - f_j\rvert}\Big).
\end{align}
Afterwards, let us define
\begin{align*}
      \tilde{\bm{A}}_{i,j}(\bm{x})=-\bigl\lvert\langle\bm{a}_i-\bm{a}_j,\bm{x}\rangle\bigr\rvert+2\nu\ln\lvert f_i - f_j\rvert, ~ \forall (i,j),~f_j\neq f_j.
\end{align*}
Then, the objective in~\eqref{eq:opt_mean_surogate_2} becomes 
\begin{align*}
    \hat{\bm{x}}
    \;=\;
    \underset{\bm{x}}{\mathrm{argmin}}
\;\sum_{i,j}\exp\Big({\frac{\tilde{\bm{A}}_{i,j}(\bm{x})}{\nu}}\Big),\quad \|\bm{x}\|_2^2\leq 1.
\end{align*}
or, equivalently, we can minimize its $\log$-transform:
\begin{align}
    \label{eq:opt_mean_surogate_3}
    \hat{\bm{x}}
    \;=\;
    \underset{\bm{x}}{\mathrm{argmin}}
    \ln\!
    \Bigl(\,
      \sum_{i,j} \exp\Big({\frac{\tilde{\bm{A}}_{i,j}(\bm{x})}{\nu}}\Big)
    \Bigr),\quad \|\bm{x}\|_2^2\leq 1.
\end{align}
Now,  we can use the following upper and lower bounds on the log-sum-exp function~\cite{calafiore2020universal}, i.e., 
\begin{subequations}
    \label{eq:upper_lower}
    \begin{align}
    \Bigg|\max_{i,j} \{\tilde{\bm{A}}_{i,j}(\bm{x})\} - \nu\ln\bigg(\sum_{i,j}\exp\Big({\frac{\tilde{\bm{A}}_{i,j}(\bm{x})}{\nu}}\Big)\bigg)\Bigg| \leq  \nu\log(|\mathcal{Y}_f|)
\end{align}
\end{subequations}
Hence, invoking the given upper and lower bound in \eqref{eq:upper_lower} for $\nu \rightarrow 0^{+}$,  the optimization problem in \eqref{eq:opt_mean_surogate_3} becomes approximately equivalent to the following min-max problem. 
\begin{align}
    \label{eq:opt_mean_surogate_4}
    \hat{\bm{x}}
    \;=\;
    \underset{\|\bm{x}\|_2^2\le 1}{\mathrm{argmin}}
    \;
    \max_{i,j}
    \Bigl\{
      -\,\tfrac{\bigl\lvert\langle\bm{a}_i-\bm{a}_j,\bm{x}\rangle\bigr\rvert}{\nu}
      \;+\;
      2\,\ln\lvert f_i - f_j\rvert
    \Bigr\}.
\end{align}
Extracting the negative sign yields an equivalent form akin to
\begin{align}
\label{eq:opt_average_max_min}
        \hat{\bm{x}}
    \;=\;
    \underset{\|\bm{x}\|_2^2\le 1}{\mathrm{argmax}}
    \;
    \min_{i,j}
    \Bigl\{
       \bigl\lvert  r_i-r_j \bigr\rvert
       \;- \tilde{\mu}_{i,j}
    \Bigr\},
\end{align}
where $\tilde{\mu}_{i,j}= 2\nu\ln\lvert f_i - f_j\rvert$ for all $(i,j)$. Similar to procedure in Section~\ref{sec:Generlzied_normal} and employing the subexpoenetial upper bound in \eqref{eq:propSubExponential}, it is straightforward to check that minimizing the maximum error $\Xi_{\rm max}$ (the maximum pairwise error probability) results in the same optimization as \eqref{eq:opt_average_max_min}. Hence, we can conclude the proof.

\bibliographystyle{IEEEtran}
\bibliography{IEEEabrv,Ref}

\end{document}